\documentclass[fleqn,usenatbib]{mnras}
\usepackage{newtxtext,newtxmath}
\usepackage[T1]{fontenc}

\usepackage{graphicx}
\usepackage{multirow, bm, IEEEtrantools}
\usepackage{tabularx}

\usepackage{acro}

\DeclareRobustCommand{\VAN}[3]{#2}
\let\VANthebibliography\thebibliography
\def\thebibliography{\DeclareRobustCommand{\VAN}[3]{##3}\VANthebibliography}

\DeclareAcronym{tov}{
  short=TOV,
  long=Tolmann-Oppenheimer-Volkoff,
}

\DeclareAcronym{sm}{
  short=SM,
  long=standard model,
}

\DeclareAcronym{ns}{
  short=NS,
  long=neutron star,
}

\DeclareAcronym{hs}{
  short=HS,
  long=hybrid star,
}

\DeclareAcronym{qcd}{
  short=QCD,
  long=quantum chromodynamics,
}

\DeclareAcronym{pqcd}{
  short=pQCD,
  long=perturbation quantum chromodynamics,
}

\DeclareAcronym{lqcd}{
  short=lQCD,
  long=lattice quantum chromodynamics,
}

\DeclareAcronym{eos}{
  short=EoS,
  long=equation of state,
}

\DeclareAcronym{nsm}{
  short=NSM,
  long=neutron star matter,
}

\DeclareAcronym{hm}{
  short=HM,
  long= hadronic matter,
}

\DeclareAcronym{ddb}{
  short=DDB,
  long=density depended couplings with Bayesian analysis,
}

\DeclareAcronym{rmf}{
  short=RMF,
  long=relativistic mean field,
}

\DeclareAcronym{nro}{
  short=NRO,
  long=non-radial oscillation,
}

\DeclareAcronym{ai}{
  short=AI,
  long=artificial intelligence,
}

\DeclareAcronym{gw}{
  short=GW,
  long=gravitational wave,
}

\DeclareAcronym{gr}{
  short=GR,
  long=general relativity,
}

\DeclareAcronym{nicer}{
  short=NICER,
  long=Neutron Star Interior Composition ExploreR,
}

\DeclareAcronym{hp}{
  short=HP,
  long=hadronic phase,
}

\DeclareAcronym{mp}{
  short=MP,
  long=mixed phase,
}

\DeclareAcronym{qp}{
  short=QP,
  long=quark phase,
}

\DeclareAcronym{njl}{
  short=NJL,
  long=Nambu--Jona-Lasinio,
}

\DeclareAcronym{ml}{
  short=ML,
  long=machine learning,
}

\DeclareAcronym{nl}{
  short=NL,
  long=non linear,
}

\DeclareAcronym{pca}{
  short=PCA,
  long=principal component analysis,
}

\DeclareAcronym{qnm}{
  short=QNM,
  long=quasi-normal mode,
}
\DeclareAcronym{dm}{
  short=DM,
  long=dark matter,
}
\DeclareAcronym{adm}{
  short=ADM,
  long= admixed dark matter,
}

\title{ Imprints of Non-Symmetric Dark Matter Halos on Magnetars: A Two-Fluid Perspective}
\author[A. Karan et al.]{Asit Karan,$^{1}$\thanks{asit22@iiserb.ac.in}
Anil Kumar,$^{2}$
Monika Sinha,$^{2}$
Ritam Mallick,$^{1}$\thanks{mallick@iiserb.ac.in}
\\
$^{1}$Department of Physics, Indian Institute of Science Education and Research, Bhopal, 462 066, India\\
$^{2}$Department of Physics, Indian Institute of Technology Jodhpur, 342030, India
}

\begin{document}
\label{firstpage}
\pagerange{\pageref{firstpage}--\pageref{lastpage}}
\maketitle

\begin{abstract}
In this study, we investigate the impact of dark matter on the structure and deformation of magnetars. We assume a perturbative approach for the magnetic field deformation and that the dark matter only interacts gravitationally with hadronic matter. Assuming that dark matter is significantly softer than hadronic matter, we find that the magnetic field can affect dark matter through the deformation of space-time. The number of stars having a dark matter halo outside the visible surface of the star increases with an increase in dark matter fraction and the stiffness of the dark matter equation of state. As the magnetic field deforms the stars from sphericity, we can have a situation where we have a non-symmetric dark matter halo outside the star. The deformation of the dark matter halo gives rise to the discrepancies in the observed period $P$ (and period derivative $\dot{P}$) and gravitational wave signatures. The observed visible surfaces predict a lower period and gravitational wave strain than that with a dark matter halo. This can have interesting observational gravitational signatures unique to magnetars having a dark matter halo.
\end{abstract}



\section{Introduction}

Observational evidence suggests that only a small fraction of the Universe's total energy density is made up of visible (baryonic) matter, accounting for approximately $\sim 5\%$, while the remaining $\sim 95\%$ consists of dark components—about $27\%$ as dark matter (DM) and $68\%$ as dark energy~\citep{Planck:2013pxb,Planck:2018vyg}. Strong evidence for the existence of DM emerged from the study of galactic rotation curves and the dynamics of galaxy clusters, which indicated the presence of unseen mass influencing gravitational behavior~\citep{Rubin:1980zd}. 
Further constraints from Big Bang nucleosynthesis and the cosmic microwave background suggest that DM is predominantly non-baryonic and must consist of a new form of matter that interacts very weakly, if at all, with standard model particles~\citep{Bertone:2004pz}.
Although the fundamental properties of DM particles remain unknown, particle physics models have imposed stringent constraints on their mass and coupling parameters~\citep{Feng:2010gw,PhysRevLett.109.021301,PhysRevLett.100.021303,PhysRevLett.131.041003,2017PhRvL.119r1301A,PhysRevLett.131.041002,PhysRevLett.127.261802}. 

There are various models of DM candidates, ranging from fermionic to bosonic particles. Among these models, the weakly interacting massive particle (WIMP) paradigm stands out due to its natural ability to produce the observed relic abundance of DM through weak-scale interactions~\citep{Nussinov:1985xr,Jungman:1995df,Gudnason:2006yj,Gudnason:2006ug,Ryttov:2008xe,Foadi:2008qv,Frandsen:2009mi,Graesser:2011wi,Gao:2011ka,Buckley:2011ye,Lewis:2011zb,Bell:2011tn,Arcadi:2017kky}. This makes WIMPs testable in terrestrial experiments and motivates their consideration in astrophysical contexts. Other popular models include Axions and sterile neutrinos \citep{PRESKILL1983127,Duffy_2009,2022SciA....8J3618C,PhysRevLett.72.17}. Axions were first proposed to solve the strong CP problem in quantum chromodynamics, but recently have gained interest as a cold DM candidate \citep{PhysRevD.80.035024,HWANG20091}. As these hypothetical particles only interact gravitationally, they are also a good candidate for DM \citep{Feng:2010gw}. However, none of the particles have been discovered yet, and there is still no consensus about a particular candidate for DM. 

Recently, compact stars such as neutron stars (NS) have emerged as astrophysical laboratories for exploring the properties of DM \citep{deLavallaz:2010wp,Sandin:2008db}. These dense stellar remnants offer unique environments to study the gravitational and potential non-gravitational interactions between DM and ordinary matter. In particular, the influence of DM on the structure, stability, and observable features of compact stars has garnered significant interest as a promising indirect probe for understanding DM properties.

As DM is more abundant than normal hadronic matter (HM), it is expected to give rise to structures as it interacts gravitationally with normal visible matter. Therefore, they can be accreted in stars and other compact objects. Interesting phenomena are expected to be observed gravitationally if DM is accumulated in stars or in compact objects. DM can be accreted in neutron stars with different mechanisms like gravitational capture via nucleon scattering \citep{PhysRevD.77.023006}, self-interacting dark matter (SIDM) capture \citep{2014JCAP...05..013G}, Bose-Einstein condensation of bosonic DM \citep{PhysRevD.40.3221}, and fermionic DM accretion \citep{PhysRevD.85.023519}. The maximum amount of DM can be captured through gravitational mechanisms involving nucleon scattering and SIDM. While gravitational capture without any interactions is possible, it is highly inefficient and contributes negligibly compared to interaction-based mechanisms. Recently, there has been a plethora of works discussing the accumulation of DM in NSs \citep{Brito:2015yga,Gleason:2022eeg,Nelson:2018xtr,Kouvaris:2010vv,Kouvaris:2015rea}. NSs are in themselves fascinating objects that are thought to harbor exotic matter at their cores \citep{Olinto:1986je,Glendenning:1997wn,Baldo:2002ju,Alford:2004pf,Drago:2005yj,Dexheimer:2009hi,Orsaria:2013hna,Bhattacharyya:2006vy,Bhattacharyya:2007dt,Prasad:2017msy,Prasad:2019kuz,Mallick:2020bdc,Annala:2019puf,Ferreira:2020evu,Han:2020adu,Prasad:2022dom,Haque:2022dsc}. As the core of NS is very dense, the matter content at its core is still not very certain. There has been work that discusses that the core of NS can have DM, which is therefore termed dark matter admixed NS (DMANS).

As the microscopic nature of DM still remains a mystery, recent work on NS has focused on analyzing NS data and constraining some of the properties of the DM. The task is not easy, and such avenues are only starting to be looked at recently. However, the recent advancement of NS observation in terms of NICER, pulsar observation, and GW data has given us hope to advance in this field. There has been work on modelling the DM as Fermionic \citep{Goldman:2013qla,Kouvaris:2015rea} and Bosonic models \citep{Karkevandi:2021ygv}. Some new and interesting models of DM, including WIMP \citep{Jungman:1995df} and Axion \citep{Duffy_2009}. As the matter properties of hadronic matter, the equation of state (EoS), in itself, is not known at those densities, and an additional assumption of DM complicates the situation. However, given some general constraints on the DM properties, it is expected that the DMANS mass-radius (M-R) will be affected by the presence of DM. There has been numerous works regarding such aspects in the last few years \citep{Ellis:2018bkr,Karkevandi:2021ygv,Leung:2011zz,Panotopoulos:2017idn,Quddus:2019ghy,Thakur_2024,Das:2020vng,Kumar:2022amh,Lourenco:2021dvh,Das:2020ptd,Dutra:2022mxl,Dengler:2021qcq,Das:2020ecp}.

If the DM resides inside the NS, the prospect of detecting it is not very great. However, things become interesting if we have some DM halo \citep{Nelson:2018xtr} outside the star, which can have gravitational interaction and even can curve the space-time (ST) around a NS, affecting its observational signatures. Things get more interesting when the halo structure is not symmetric outside the NS and has some asymmetry. One of the most common and natural causes of asymmetry in NS ST can arise either due to rotation or due to the magnetic field. However, to affect the ST around an NS significantly, the magnetic field has to be ultra-strong. This can only happen in the case of magnetars.

Magnetars are NS that have an ultra-strong magnetic field at their surface. A few tens of magnetars have been detected in the last few decades and are always associated with anomalous X-ray pulsars or soft gamma-ray repeaters. The surface magnetic fields of magnetars are deduced to be of the order of a few $10^{14}-10^{15}$ G \citep{Duncan:1992hi,Kouveliotou:1998ze}. The magnetic field inside the NS cannot be directly deduced, but various models have predicted that it can even be of the order of $10^{18}$ G \citep{Ferrer:2010wz,Dexheimer:2011pz}. Such a high magnetic field deforms the magnetars and also the ST around it. It can even affect the EoS; however, its effect is negligible compared to the curvature effect \citep{Chakrabarty:1997ef,Sinha:2010fm,Mallick:2011dd,Dexheimer:2011pz,Dexheimer:2012mk}.

Modelling a magnetar's magnetic field is non-trivial. However, to keep our model simple and traceable, we adopt a technique that is perturbative in nature. Incorporating the magnetic field in the energy-momentum tensor, we solve the Einstein equation perturbatively. To incorporate the DM effect, we also adopt a two-fluid approach, where the normal HM and the DM only interact gravitationally. Although the magnetic field does not affect the DM directly, it would be interesting to note that the deformed curvature of NS ST implicitly affects the DM and gives rise to some of the most interesting results.

The aim of this project is precisely to solve the NS structure in a two-fluid approach, having HM and DM, and observing the effect of the magnetic field on them. To have a considerable effect of the magnetic field on the DMANS structure, the magnetic field needs to be sufficiently strong, which is only available in magnetars. We are not much interested in determining the particle nature of DM, but rather in how one can detect dark matter halos around DMANS, especially in the case of magnetars.
The paper is arranged as follows: In the next section (section \ref{Theory}), we discuss the formalism applied in the work. In section \ref{result}, we present our results, and finally, in section \ref{sum_disc}, we summarize and conclude our work. Throughout this work, we use natural units where \( c = \hbar = G = 1 \).

\section{Theoretical Formalism}\label{Theory}
In this section, we present the fundamental equations relevant to our study. We begin with the EoS for HM and DM in the context of dark matter-admixed magnetized neutron stars. This is followed by formulating the two-fluid hydrostatic equilibrium conditions for DMANSs and, subsequently, the set of equations governing their deformation due to magnetic fields and the magnetic field profile within the DMANSs.

\subsection{Hadronic matter EoS}
We assume that HM is composed of nucleons (N) and leptons (l) such as electron (e) and muon ($\mu$). We use a well-established covariant density functional (CDF) model for the EoS of HM, which includes the density-dependent parametrization. 
\subsubsection*{CDF model} In this model, the scalar interaction between nucleons is mediated via $\sigma$ meson, and vector interactions via $\omega$ and $\rho$ mesons. The Lagrangian density of HM with the CDF model can be given as \citep{RING1996193,Glendenning:1997wn}
\begin{align}
\mathcal{L} & = \sum_{N} \bar{\psi}_N(i\gamma_{\mu} D^{\mu} - m^{*}_N) \psi_N + \sum_{l} \bar{\psi}_l (i\gamma_{\mu} \partial^{\mu} - m_l)\psi_l \nonumber\\ 
 & -  \frac{1}{4}\omega_{\mu\nu}\omega^{\mu\nu} + \frac{1}{2}m_{\omega}^2\omega_{\mu}\omega^{\mu} - \frac{1}{4}\boldsymbol{\rho}_{\mu\nu} \cdot \boldsymbol{\rho}^{\mu\nu} + \frac{1}{2}m_{\rho}^2\boldsymbol{\rho}_{\mu} \cdot \boldsymbol{\rho}^{\mu}\label{eqn:lag_NM} 
\end{align}
Here, $\psi_N$ and $\psi_l$ denote the field of nucleons and leptons, respectively, with their masses. The covariant derivative is $D_\mu = \partial_\mu + ig_{\omega N} \omega_\mu + ig_{\rho N} \boldsymbol{\tau}_{N3} \cdot \boldsymbol{\rho}_{\mu}$ and the effective mass of nucleon is $m_{N}^* = m_N - g_{\sigma N}\sigma$. In this equation, vector fields are $\omega_{\mu \nu}  = \partial_{\mu}\omega_{\nu} - \partial_{\nu}\omega_{\mu}$ and $\boldsymbol{\rho}_{\mu \nu}  = \partial_{\nu}
\boldsymbol{\rho}_{\mu} - \partial_{\mu}\boldsymbol{\rho}_{\nu}$. Considering density-dependent parametrization, the coupling strength of nucleon with the $i$-th meson can be given as \citep{TYPEL1999331,PhysRevC.66.024306}
\begin{equation}
g_{i N}(n)= g_{i N}(n_{0}) f_i(x) \quad \quad \text{for }i=\sigma,\omega
\end{equation}
where the function is given by
\begin{equation}\label{eqn.func}
f_i(x)= a_i \frac{1+b_i (x+d_i)^2}{1+c_i (x +d_i)^2}
\end{equation}
where $x=n/n_0$ and $a_i$, $b_i$, $c_i$, $d_i$ are constants, describing properties at saturation density ($n_0$).
But, for the $\rho$-meson, the density-dependent coupling parameter is 
\begin{equation}
g_{\rho N}(n)= g_{\rho N}(n_{0}) e^{-a_{\rho}(x-1)}
\end{equation}
The chemical potential of a nucleon can be obtained as
\begin{equation}
\mu_N = \sqrt{{p_{f_N}}^2 + {m_N^*}^2} + g_{\omega N}\omega_0 + g_{\rho N} \boldsymbol{\tau}_{N3}{\rho}_{03} + \Sigma^r
\end{equation}
Where $\Sigma^r$ is the rearrangement term that arises due to the density-dependent parametrization given by \citep{PhysRevC.64.025804}
\begin{equation}
\begin{aligned}
\Sigma^{r} & = \sum_{N} \left[ \frac{\partial g_{\omega N}}{\partial n}\omega_{0}n_{N} - \frac{\partial g_{\sigma N}}{\partial n} \sigma n_{N}^s + \frac{\partial g_{\rho N}}{\partial n} \rho_{03} \boldsymbol{\tau}_{N3} n_{N} \right]
\end{aligned}
\end{equation}
In this rearrangement, term $n$ denotes the number density and $n^s$  denotes the scalar number density. The energy density of HM with this model is
\begin{align}
\varepsilon_{\mathrm{hm}} &= \frac{1}{\pi^2} \bigg(\sum_N \int_{0}^{p_{f_N}}k^2\sqrt{k^2+{m_N^*}^2}dk \nonumber\\ 
&+\sum_l \int_{0}^{p_{f_l}}k^2\sqrt{k^2+{m_l}^2}dk \bigg) + \frac{1}{2}m_{\sigma}^2 \sigma^{2} \nonumber\\ &+ \frac{1}{2} m_{\omega}^2 \omega_{0}^2 + \frac{1}{2}m_{\rho}^2 \rho_{03}^2 
\end{align}
Here, $p_f$ is the Fermi momentum, and pressure can be evaluated by the relation 
\begin{equation} 
 P_{\mathrm{hm}} = \sum_{i=N,l} \mu_in_i - \varepsilon_{\mathrm{hm}}
\end{equation}
We opt to choose DDME2 parametrization \citep{PhysRevC.71.024312} in this work because it goes well with recent astrophysical observations, also \citep{10.1093/mnras/stab2327}.

\subsection{Dark matter EoS}
Given the incomplete understanding of DM, DM particles may be either fermions \citep{PhysRevC.89.025803,Thakur_2024,PhysRevD.89.015010} or bosons \citep{PhysRevD.87.055012,PhysRevD.105.023001} in compact stars. DM particles can span a wide mass range. We assume they are free fermions and disregard possible self-interactions for the sake of simplification. The Lagrangian density of free fermionic DM can be given as \citep{PhysRevD.103.043009}
\begin{align}
    \mathcal{L}_{\mathrm{dm}} = &\bar{\psi_D}[i\gamma_\mu\partial^\mu - m_D]\psi_D
\end{align}
EoS of fermionic DM can be evaluated as
\begin{align}
    \varepsilon_{\mathrm{dm}} &= \frac{1}{\pi^2} \int_0^{p_{f_D}}k^2\sqrt{k^2 + {m_D}^2}dk \\
    P_{\mathrm{dm}} &= \frac{1}{3\pi^2} \int_0^{p_{f_D}}\frac{k^4}{\sqrt{k^2 + {m_D}^2}}dk
\end{align}
Here, $p_{f_D}$ is the Fermi momentum of the DM particle, $m_D$ is the bare mass and the number density is
\begin{align}
    n_D = \frac{{p_{f_D}}^3}{3\pi^2}
\end{align}

In principle, one can have more complicated interaction terms in the Lagrangian; however, the aim of this paper is not to determine the actual nature of DM but to study DM halos and their possible detection in a strong magnetic regime. Therefore, we keep our DM EoS simple to easily analyze the interesting aspects detailed above. 

\subsection{Two fluid TOV}
The structure equations describing a non-rotating, spherically symmetric star composed of a perfect single fluid are given by the Tolman–Oppenheimer–Volkoff (TOV) equations \citep{Tolman:1939jz,Oppenheimer:1939ne}.
However, when considering a multicomponent fluid, such as a DMANS, the TOV equations are modified to the two-fluid TOV form, where one fluid, DM, interacts with the other, HM, only through gravity \citep{Ciarcelluti:2010ji}. In the absence of a magnetic field, these objects are spherically symmetric. The spacetime around a static, non-rotating, spherically symmetric DMANS is described by the metric:
\begin{IEEEeqnarray}{rCl}
ds^2 &=& -e^{2\nu(r)}(dt)^2 + e^{2\lambda(r)}(dr)^2 \nonumber\\
&&\hspace{4.8em} + r^2\left((d\theta)^2 + \sin^2\theta\, (d\phi)^2\right) \label{tov.metric}
\end{IEEEeqnarray}
where \( \nu(r) \) and \( \lambda(r) \) are the metric potentials.

Using the background metric (\ref{tov.metric}), and solving the Einstein equations in the presence of two distinct fluids with conserved energy-momentum tensors for HM, \( T_{\mu\nu}^{\mathrm{(hm)}} \), and DM, \( T_{\mu\nu}^{\mathrm{(dm)}} \), the two-fluid TOV equations are given by \citep{Sandin:2008db,Ciarcelluti:2010ji,Das:2020ecp}:

\begin{align}
\frac{dP_{\mathrm{hm}}}{dr} &= -\frac{(P_{\mathrm{hm}} + \varepsilon_{\mathrm{hm}})\left\{m + 4\pi r^3(P_{\mathrm{hm}} + P_{\mathrm{dm}})\right\}}{r(r - 2m)} \label{2tov.psr1}\\
\frac{dP_{\mathrm{dm}}}{dr} &= -\frac{(P_{\mathrm{dm}} + \varepsilon_{\mathrm{dm}})\left\{m + 4\pi r^3(P_{\mathrm{hm}} + P_{\mathrm{dm}})\right\}}{r(r - 2m)} \label{2tov.psr2}
\end{align}

with
\begin{IEEEeqnarray}{rCl}
\frac{dm(r)}{dr} &=& 4\pi (\varepsilon_{\mathrm{hm}} + \varepsilon_{\mathrm{dm}})r^2 \label{2tov.mass}
\end{IEEEeqnarray}
The total gravitational mass enclosed within a radius \( r \) is defined as \( m(r) = m_{\mathrm{hm}}(r) + m_{\mathrm{dm}}(r) \), where \( m_{\mathrm{hm}}(r) \) and \( m_{\mathrm{dm}}(r) \) are the mass contributions from HM and DM, respectively. Here, \( P_{\mathrm{hm}} \) and \( P_{\mathrm{dm}} \) denote the pressures, while \( \varepsilon_{\mathrm{hm}} \) and \( \varepsilon_{\mathrm{dm}} \) represent the energy densities of the HM and DM components, respectively.

The metric potentials \( \nu(r) \) and \( \lambda(r) \) in the two-fluid formalism are given by:
\begin{IEEEeqnarray}{rCl}
    \frac{d\nu(r)}{dr} &=& \frac{\left\{m + 4\pi r^3(P_{\mathrm{hm}} + P_{\mathrm{dm}})\right\}}{r(r - 2m)}\label{2tov.nu}\\
    \lambda(r) &=& -\frac{1}{2}\ln{\left\{1 - \frac{2(m_{\mathrm{hm}} + m_{\mathrm{dm}})}{r}\right\}}\label{2tov.lamda}
\end{IEEEeqnarray}
These coupled differential equations - Eqs.~(\ref{2tov.psr1}), (\ref{2tov.psr2}), (\ref{2tov.mass}), (\ref{2tov.nu}), and (\ref{2tov.lamda}) can be solved numerically with appropriate boundary conditions. The surfaces of the HM and DM are defined by the radii at which their respective pressure vanish: \( P_{\mathrm{hm}}(R_{\mathrm{hm}}) = 0 \) defines the surface of the HM at radius \( R_{\mathrm{hm}} \), while \( P_{\mathrm{dm}}(R_{\mathrm{dm}}) = 0 \) defines the surface of the DM at radius \( R_{\mathrm{dm}} \). The total radius \( R \) of the star is given by the maximum of the two, i.e., \( R = \max(R_{\mathrm{hm}}, R_{\mathrm{dm}}) \).
Depending on the choice of the energy density for HM  and DM EOSs, different structural configurations can emerge. For instance, a DM core configuration corresponds to \( R_{\mathrm{hm}} > R_{\mathrm{dm}} \), a DM halo structure corresponds to \( R_{\mathrm{dm}} > R_{\mathrm{hm}} \), and a mixed configuration arises when \( R_{\mathrm{hm}} = R_{\mathrm{dm}} \).
The amount of DM inside a NS is defined by the \textit{DM fraction} as
\begin{IEEEeqnarray}{rCl}
    \mathcal{F}_{\mathrm{dm}} &=& \frac{M_{\mathrm{dm}}}{M_{\mathrm{hm}} + M_{\mathrm{dm}}}
\end{IEEEeqnarray}
where \( M_{\mathrm{dm}} \) and \( M_{\mathrm{hm}} \) are the masses of DM and HM, respectively. One can regulate the amount of DM inside a DMANS by varying the fraction or the bare mass ($m_D$) of the DM.

\subsection{Magnetic perturbation in general relativity due to magnetic field}
In this section, we consider the deformation of the DMANS due to the presence of a magnetic field and treat this deformation perturbatively, analogous to the perturbative approaches used for rotating neutron stars by Hartle \citep{Hartle:1967he}.
Under the influence of the magnetic field, the star becomes distorted, leading to perturbations in the pressure, energy density, and number density, denoted by $\delta P$, $\delta \varepsilon$, and $\delta n_b$, respectively. These perturbations, in turn, modify the energy-momentum tensor by an amount $\delta T_{\mu\nu}$, so that the total energy-momentum tensor becomes:
\begin{IEEEeqnarray}{rCl}
     T_{\mu\nu} = T^0_{\mu\nu} + \delta T_{\mu\nu}
\end{IEEEeqnarray}
Where,
\begin{IEEEeqnarray}{rCl}
     T^0_{\mu\nu} &=& T^{0(hm)}_{\mu\nu} + T^{0(dm)}_{\mu\nu} \nonumber \\
                  &=&\left(\varepsilon + P\right)u_{\mu} u_{\nu} + Pg_{\mu\nu},\\[5pt]
     \delta T_{\mu\nu} &=& \left(\delta \varepsilon + \delta P\right)u_{\mu} u_{\nu} + \delta Pg_{\mu\nu}                 
\end{IEEEeqnarray}
Here, the quantity $T^{0}_{\mu\nu}$ denotes the perfect fluid energy-momentum tensor for the two-fluid configuration, consisting of the energy-momentum tensor for HM, $T^{0(hm)}_{\mu\nu}$, and that for DM, $T^{0(dm)}_{\mu\nu}$. The total energy density of the matter is given by $\varepsilon = \varepsilon_{\mathrm{hm}} + \varepsilon_{\mathrm{dm}}$, and the total pressure by $P = P_{\mathrm{hm}} + P_{\mathrm{dm}}$.
Assuming the poloidal magnetic configuration, one can write the pressure perturbation $\delta P$ in lowest order in terms of spherical harmonics as,
\begin{IEEEeqnarray}{rCl}
    \delta P =  \left(p_0 + p_2 P_2(\cos{\theta})\right)
\end{IEEEeqnarray}
Where $p_0$ and $p_2$ are the monopole and quadrupole contributions of magnetic pressure. The second-order Legendre polynomial is $P_2(\cos{\theta})$ define as,
\begin{IEEEeqnarray}{rCl}
    P_2(\cos{\theta}) =  \frac{1}{2}\left(3\cos^2{\theta} - 1\right)
\end{IEEEeqnarray}
The line element for the non-rotating magnetically deformed star can be written as:
\begin{IEEEeqnarray}{rCl}
ds^2 &=& -e^{2\nu(r,\theta,\varepsilon_B)}(dt)^2 + e^{2\lambda(r,\theta,\varepsilon_B)}(dr)^2 \nonumber\\
     && \quad + e^{2\alpha(r,\theta,\varepsilon_B)}(d\theta)^2 + e^{2\beta(r,\theta,\varepsilon_B)}(d\phi)^2 \label{metric}
\end{IEEEeqnarray}
The metric function $\nu$, $\lambda$, $\alpha$, and $\beta$ in the perturbative approach can be expanded up to the second order in the star magnetic energy density ($\varepsilon_B$) as 
\begin{IEEEeqnarray}{rCl}
    e^{2\nu(r,\theta,\varepsilon_B)} &=& e^{2\nu(r)}\left\{1 + 2\left[h_{0}(r,\varepsilon_B) \right.\right. \nonumber\\
                         &&\hspace{3.5em}\left.\left. +\; h_{2}(r,\varepsilon_B)P_{2}(\cos{\theta})\right]\right\}\\[5pt]                   
     e^{2\lambda(r,\theta,\varepsilon_B)} &=& e^{2\lambda(r)}\left\{1 + \frac{e^{2\lambda}}{r}\left[m_{0}(r,\varepsilon_B) \right.\right. \nonumber\\
                         && \hspace{3.7em}\left.\left. +\; m_{2}(r,\varepsilon_B)P_{2}(\cos{\theta})\right]\right\}\\[5pt]
    e^{2\alpha(r,\theta,\varepsilon_B)} &=& r^{2}\left\{1 + 2w_{2}(r,\varepsilon_B)P_{2}(\cos{\theta})\right\} \\ [5pt]   
    e^{2\beta(r,\theta,\varepsilon_B)} &=& r^{2}\sin^{2}{\theta}\left\{1 + 2w_{2}(r,\varepsilon_B)P_{2}(\cos{\theta})\right\}     
\end{IEEEeqnarray}
Where the functions \( h_0 \) and \( m_0 \) represent the monopole perturbations of the metric, while \( h_2 \), \( m_2 \), and \( w_2 \) correspond to the quadrupole perturbations.

Now, since the metric is perturbed, the Einstein tensor \( G_{\mu\nu} \) also undergoes a corresponding perturbation. This perturbation, denoted by \( \delta G_{\mu\nu} \), is expressed in terms of the functions \( h_0 \), \( m_0 \), \( h_2 \), \( m_2 \), and \( w_2 \).
Using the background metric~\eqref{metric}, the equilibrium conditions for non-rotating magnetized DMANSs are obtained by solving the unperturbed part of the Einstein equations. This leads to the two-fluid TOV equations, given by~\eqref{2tov.psr1}, \eqref{2tov.psr2}, and~\eqref{2tov.mass}.
The differential equations for the monopole perturbation functions are obtained by solving the perturbed part of the Einstein equations, considering only the monopole (\( \ell = 0 \)) terms, and are given by \citep{Schramm:2013ipa}:
\begin{IEEEeqnarray}{rCl}
    \frac{dm_0}{dr} &=& 4\pi r^2 p_0 \label{perturb.mass}\\[5pt]
    \frac{dh_0}{dr} &=& 4\pi r e^{2\lambda} p_0 + \frac{d\nu}{dr}  \frac{e^{2\lambda}m_0 }{ r} +  \frac{e^{2\lambda}m_0}{ r^2} \label{perturb.h0} \\[5pt]
    \frac{dp_0}{dr} &=& -\frac{d\nu}{dr} p_0 - (\varepsilon + P) \frac{dh_0}{dr}\label{perturb.p0}
\end{IEEEeqnarray}
While solving the quadrupole ($l=2$) terms in the Einstein equations, we obtain the differential equations for the quadrupole perturbation functions, which determine the shape of non-rotating magnetized DMANSs, given by:
\begin{IEEEeqnarray}{rCl}
     \frac{d\nu}{dr}\frac{dw_2}{dr}  &=& 8\pi r (\varepsilon + P)e^{2\lambda} p_2 \nonumber\\
     &&\hspace{2.6em} + \left(6e^{2\lambda} - 2\right) \frac{h_2}{r} + 4 e^{2\lambda} \frac{w_2}{r} \label{perturb.w2}\\ [5pt]
     \frac{dh_2}{dr} &=& -\frac{d\nu}{dr} h_2 - \frac{dw_2}{dr} \label{perturb.h2}\\[5pt]
     \frac{dp_2}{dr} &=& -\frac{d\nu}{dr} p_2 - (\varepsilon + P) \frac{dh_2}{dr}\label{perturb.p2}
\end{IEEEeqnarray}
Where $\lambda$ and $\nu$ are the unperturbed metric potentials given by (\ref{2tov.lamda}) and (\ref{2tov.nu}).
We solve the perturbative equations (\ref{perturb.mass}), (\ref{perturb.h0}), (\ref{perturb.p0}), (\ref{perturb.w2}), (\ref{perturb.h2}), and (\ref{perturb.p2}) with the following boundary conditions: at the center of the star, all the metric perturbations vanish, i.e., \( h_0(0) = 0 \), \( m_0(0) = 0 \), \( w_2(0) = 0 \), and \( h_2(0) = 0 \). For the magnetic pressure perturbations, the boundary conditions are defined as \( p_0(0) = \frac{B_{c}^2}{24\pi} \) and \( p_2(0) = -\frac{B_c^2}{6\pi} \), where \( B_c \) denotes the strength of the magnetic field at the center of the star.
By solving Eq.~(\ref{perturb.mass}), we obtain \( m_0 \), which represents the change in the gravitational mass due to the presence of a magnetic field. The total mass of the magnetized DMANSs is defined as 
\begin{IEEEeqnarray}{rCl}
    M_T &=& M_0 + m_0
\end{IEEEeqnarray}
where \( M_0 \) is the mass of the unmagnetized NS (mass of the DMANS), which, in our case, corresponds to the gravitational mass obtained by solving the two-fluid TOV equation~(\ref{2tov.mass}).

The anisotropic pressure generated by the magnetic field not only increases the gravitational mass but also deforms the star. Specifically, the magnetic pressure is enhanced along the equatorial direction and reduced along the polar direction, leading to an oblate stellar shape. The equatorial and polar radii of magnetized DMANSs are given by \citep{Schramm:2013ipa}
\begin{IEEEeqnarray}{rCl}
    r_e &=& r + \eta_0(r) - \frac{1}{2}\left[\eta_2(r) + r w_2(r)\right], \\
    r_p &=& r + \eta_0(r) + \left[\eta_2(r) + r w_2(r)\right],
\end{IEEEeqnarray}
where \( r \) is the radius of the unmagnetized, spherically symmetric NS. The functions \( \eta_0(r) \) and \( \eta_2(r) \) are defined as
\begin{IEEEeqnarray}{rCl}
    \eta_0(r) &=& \frac{r(r-2m(r))}{4\pi r^3 P(r) + m(r)} p_0(r), \\
    \eta_2(r) &=& \frac{r(r-2m(r))}{4\pi r^3 P(r) + m(r)} p_2(r),
\end{IEEEeqnarray}
where \( m(r) \) is the enclosed gravitational mass, \( P(r) \) is the total matter pressure , and \( p_0(r) \) and \( p_2(r) \) are the magnetic pressure perturbations.
To compute the metric perturbations and determine the mass and deformation of the star, a magnetic field profile is required, which is discussed next.

\subsection{Magnetic Field Profile}\label{mag.profile}
In this study, we consider a number density-dependent magnetic field configuration throughout the star, given by \citep{Chakrabarty:1997ef,Bandyopadhyay:1998qs,Rabhi:2009ih,Ryu:2010zzb,Menezes:2014aka,Schramm:2013ipa}:
\begin{IEEEeqnarray}{rCl}
    B(n_b) = B_s + B_0\left[1 - e^{-\sigma\left(\frac{n_b}{n_0}\right)^{\xi}}\right] \label{mag.prof}
\end{IEEEeqnarray}
where \( n_b \) is the baryon number density and \( n_0 = 0.15~\mathrm{fm}^{-3} \) is the nuclear saturation density.  
This model generates a magnetic field that is nonuniform throughout the star. The magnetic field strength at the center differs by $2-3$ orders of magnitude from that at the surface. The parameters \( \sigma \) and \( \xi \) control how rapidly the central magnetic field \( B_0 \) decays to the surface magnetic field \( B_s \). Observationally, the surface magnetic field is found to lie in the range \( 10^{14} - 10^{15}~\mathrm{G} \) \citep{Duncan:1992hi,Kouveliotou:1998ze,Ibrahim:2003ev,Dong:2013hta}, while the central magnetic field strength can reach up to few times \( 10^{18}~\mathrm{G} \) \citep{Ferrer:2010wz,Dexheimer:2011pz}.  
In this work, we set \( \sigma = 0.01 \) and \( \xi = 2 \), and fix the surface magnetic field at \( B_s = 10^{15}~\mathrm{G} \) and the asymptotic magnetic field at \( B_0 = 4\times10^{18}~\mathrm{G} \).  
With the chosen parameters, the magnetic field obtained from Eq.~(\ref{mag.prof}) remains below \( 1.75 \times 10^{18}~\mathrm{G} \) throughout the star, even at $n_b$ corresponding to the maximum mass configuration. A choice of surface field of $10^{16}$ G would not affect our results significantly; however, a smaller value of the magnetic field at the center of the star would not have appreciable deformation due to the magnetic field.

\section{Results}\label{result}

\begin{figure}
    \centering
    \includegraphics[scale=0.42]{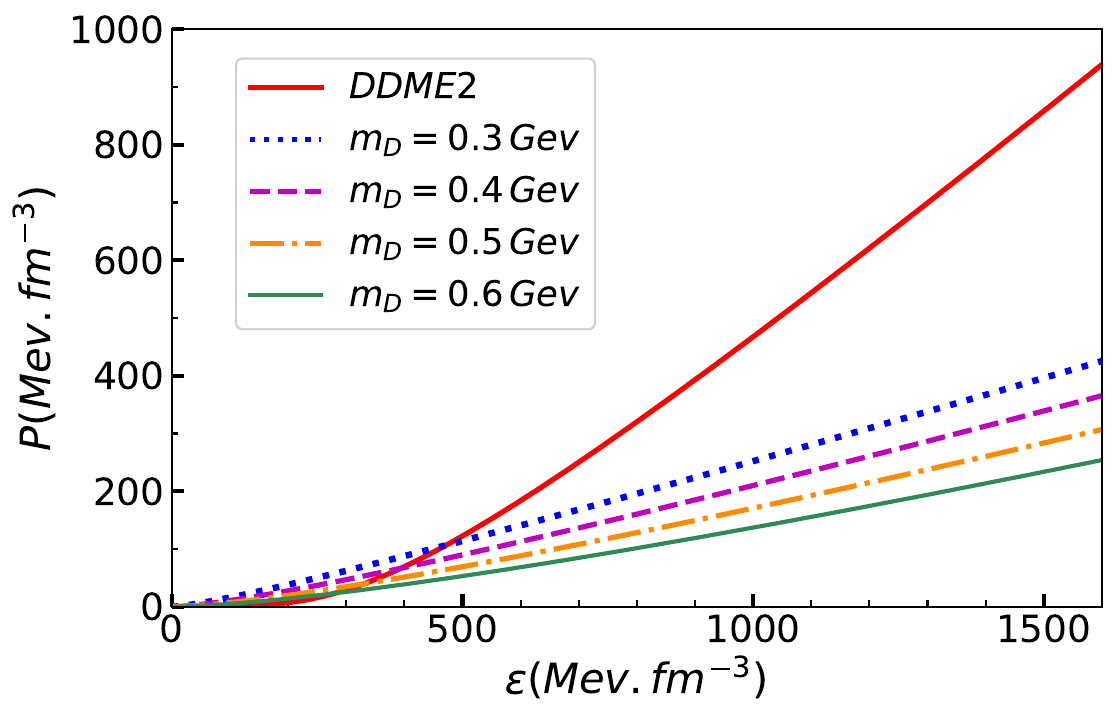}
    \includegraphics[scale=0.42]{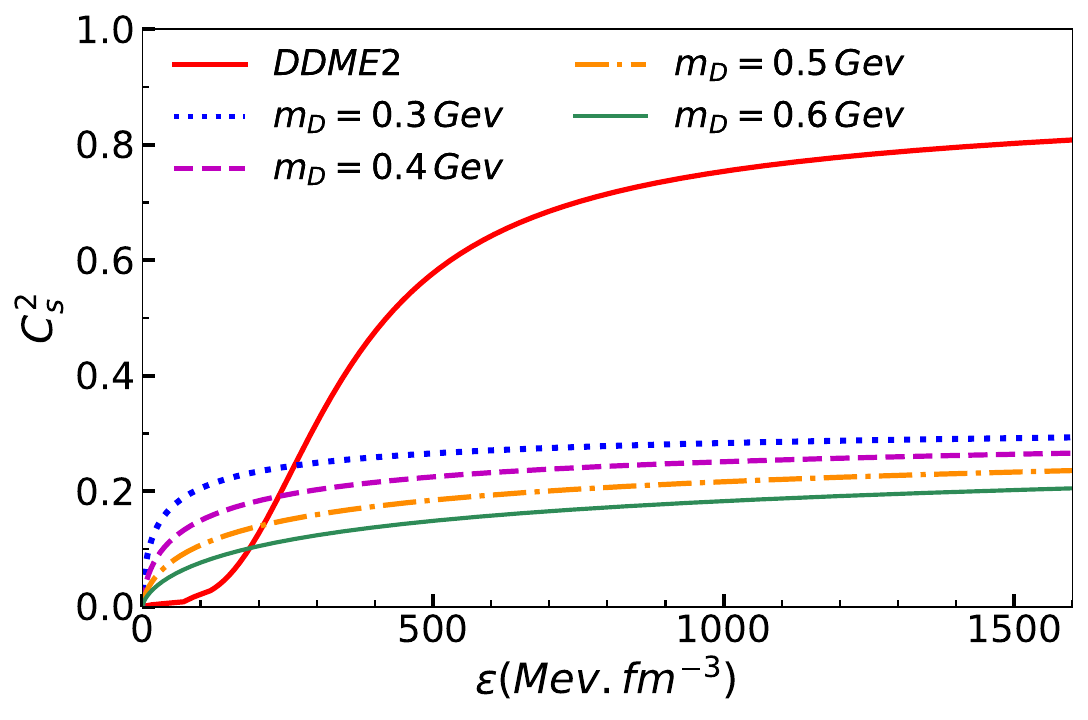}
    \caption{(\textbf{Upper}) The EoS for hadronic matter constructed using the DDME2 parametrization (red solid line), along with the EoS for dark matter with different dark matter bare masses ($m_D$) used in this work. (\textbf{Lower}) The square of the speed of sound ($C_s^2$) for different dark matter bare masses ($m_D$) and for hadronic matter as a function of energy density ($\varepsilon$). }
    \label{eos}
\end{figure}

One of the deciding factors of the structure of the DMANS is the relative EoS of HM and DM. It is usually assumed that the DM is softer than the HM. However, it depends mostly on the parameters of the DM EoS. In this treatment, we assumed that the DM is softer than the HM EoS and it has only one variable, the bare mass of dark matter $m_D$. We model the EoS of the HM with DDME2 parametrization. It is to be understood that the details of the EoS do not change the qualitative aspect of our results as long as the overall characteristics of the HM and DM EoS are similar to those shown in the upper panel of Fig. \ref{eos}. It is observed that beyond the density $\sim 300$ MeV fm$^{-3}$, the HM EoS is significantly stiffer than the DM EoS. Also, the parameter $m_D$ controls the stiffness of the DM EoS, and as $m_D$ increases, the EoS becomes softer. For this work, we have assumed the bare mass ($m_D$) to lie between $300$ and $600$ MeV. A larger bare mass makes the DM EoS so soft that it has almost a negligible effect on the star.
In the lower panel of Fig. \ref{eos}, the square of the speed of sound is plotted as a function of energy density. This clearly shows that the stiffness of the HM EoS at low energy density is smaller than the DM EoS, and beyond that, it is much stiffer. These features of the EoS at low and high density have a significant impact on the results which would follow subsequently. 

\begin{figure}
    \centering
    \includegraphics[scale=0.45]{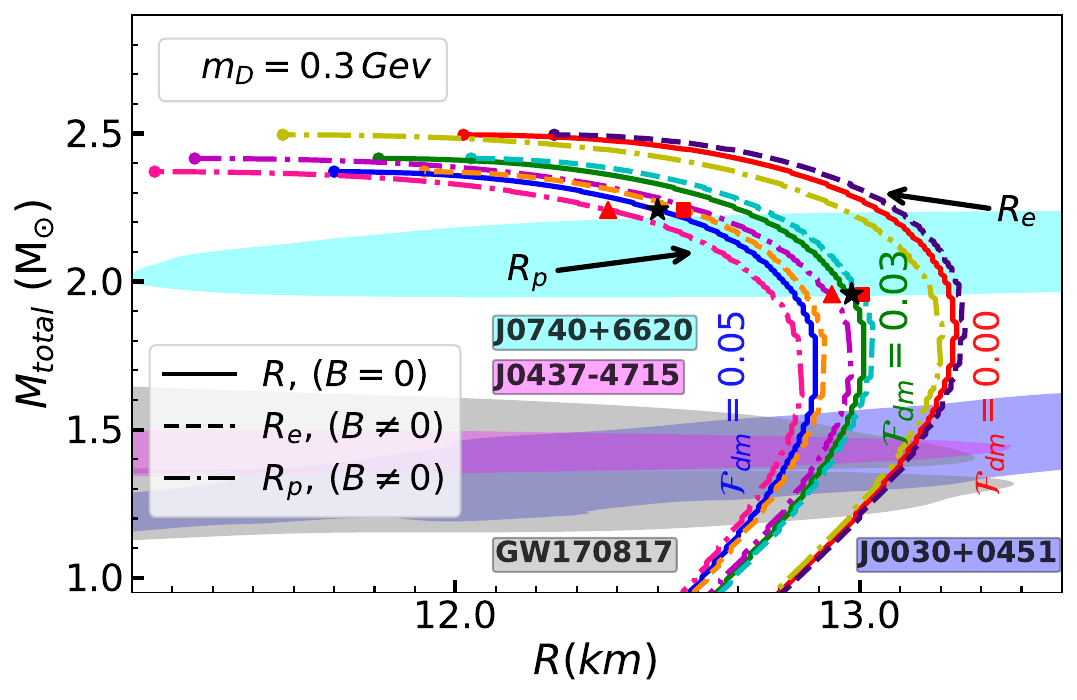}
    \caption{ The mass-radius (M-R) relation for magnetized DMANSs with a fixed bare mass ($m_D = 0.3$ GeV), shown for different dark matter fractions ($\mathcal{F}_{dm}$ = 0.00, 0.03, 0.05). The total mass is given by, $M_{total}=M_{HM} +M_{DM}+m_0$. The solid line represents a non-magnetized NS (B = 0), while the dot-dashed and dashed lines represent the polar ($R_p$) and equatorial ($R_e$) radii of magnetized DMANSs, respectively. The star symbol in the plots indicates the point where the DM and HM share the same radius ($R_{HM}=R_{DM}$); to the left of this point, DM resides in the core ($R_{HM}>R_{DM}$), and to the right, it forms a halo around the star ($R_{HM} < R_{DM}$). The red triangle and square mark the points where the polar and equatorial radii of DM and HM are equal ($R_{p,HM}=R_{p,DM}$ and $R_{e,HM}=R_{e,DM}$). On the left, DM is confined within HM along the polar direction, while on the right, it extends beyond it. The same interpretation applies to the equatorial direction. The dot markers indicate the maximum mass configuration. The gray contour represents the M–R constraints derived from the gravitational wave event GW170817 \citep{LIGOScientific:2017vwq,LIGOScientific:2018hze}, while the blue, magenta and cyan contours correspond to the M–R constraints obtained from NICER measurements of PSR J0030+0451 \citep{Riley:2019yda}, PSR J0437--4715 \citep{Choudhury:2024xbk} and PSR J0740+6620 \citep{Riley:2021pdl}, respectively.}
    \label{mr}
\end{figure}

To solve the TOV equation with a two-fluid approach, we need to have a definite fraction of DM ($\mathcal{F}_{\mathrm{dm}}$) present in the star. Therefore, for this work, $\mathcal{F}_{\mathrm{dm}}$ is also a variable. We limit the total dark matter mass fraction to 5\%, consistent with existing constraints on DM accumulation in neutron stars via accretion or inheritance from DM-rich progenitor environments \citep{Bell:2020jou,Ivanytskyi:2019wxd,Sagun:2021oml}. Asymmetric DM can be captured through scattering mechanisms in the early stages of neutron star evolution. To avoid inducing a collapse of the neutron star into a black hole, we considered a suitable range of DM mass \citep{PhysRevD.85.023519}. The M-R curve plotted in Fig. \ref{mr} shows how the magnetic field and DM impact the star structure for a two-fluid approach. In the M-R analysis, we define the stellar radius as the observable radius of HM, as electromagnetic observations used to infer the radius are insensitive to any surrounding DM halo \citep{Shakeri:2022dwg}. For stars with almost no DM, the magnetic field deforms the star, and the equatorial and polar radius differs. This difference is insignificant if the magnetic field at the center is less than $10^{17}$ G. For our analysis we have assumed that the magnetic field at the surface is fixed at $10^{15}$ G and at the center it is around $10^{17}$ to $10^{18}$ G. It does vary with the central energy density of the star as described in Section (\ref{mag.profile}). For low mass stars, the magnetic field is low at the center, and for massive stars, it is high ($B_c$ for a 2 solar mass star is around $1.5\times 10^{18}$ G). We present various astrophysical observational constraints in a single figure. The cyan-colored band represents the X-ray data from the Neutron Star Interior Composition Explorer (NICER) for the pulsar PSR~J0740+6620 \citep{Riley:2021pdl}, with an observed mass of \(2.14^{+0.10}_{-0.09}\,M_\odot\) at a 68\% confidence interval.
 We also include the millisecond pulsar PSR~J0030+0451 \citep{Riley:2019yda}, as reported by NICER, shown in the blue-colored region with a 68\% confidence level. The magenta-colored band represents the observational contour for the millisecond pulsar PSR J0437–4715 \citep{Choudhury:2024xbk} as measured by NICER. In addition to the NICER data, we incorporate the LIGO/Virgo gravitational wave observations of GW170817 \citep{LIGOScientific:2017vwq, LIGOScientific:2018hze}, with the grey region indicating the 68\% confidence level.

\begin{figure}
    \centering
    \includegraphics[scale=0.45]{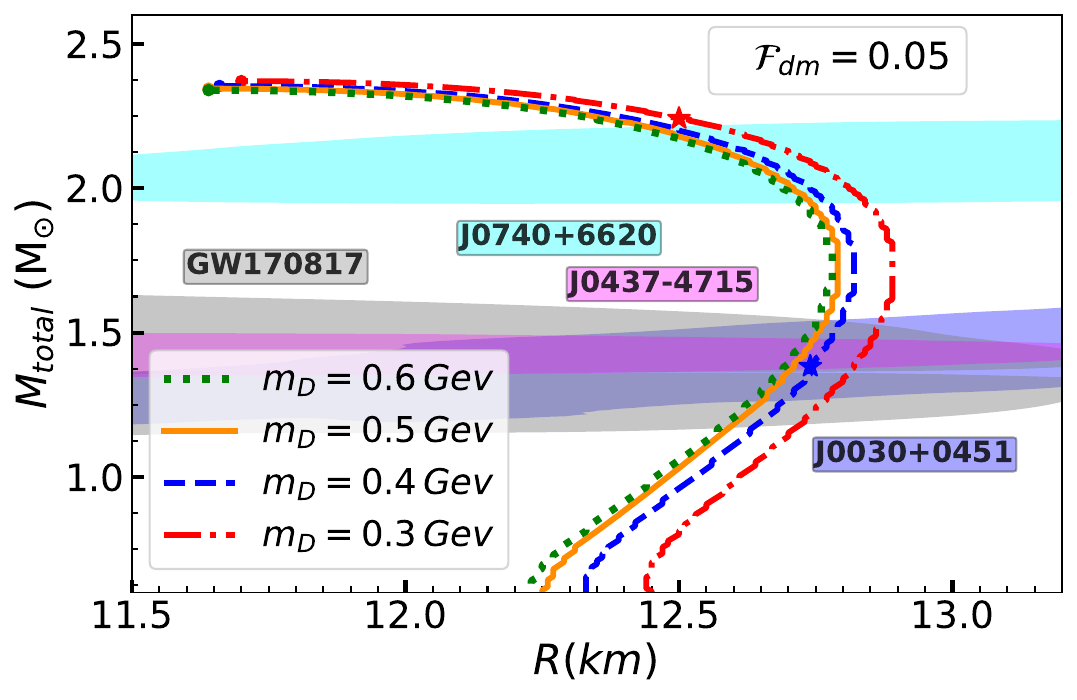}
    \caption{ M–R relation for DMANSs (B = 0) with varying $m_D$, shown for a fixed dark matter fraction $\mathcal{F}_{\mathrm{dm}} = 0.05$. The star markers indicate representative configurations: in the upper branch, the neutron stars harbour a dark matter core, whereas in the lower branch, the dark matter forms a halo. The same observational M-R contours as in Fig.\ref{mr} are included for comparison.}
    \label{mr_dm}
\end{figure}

It is seen that as $\mathcal{F}_{\mathrm{dm}}$ increases, the M-R curve shifts towards the left, indicating more compact stars. This is expected, as the EoS of DM is softer at high densities, thereby softening the overall EoS and reducing the maximum mass value. Also, it is seen that as $\mathcal{F}_{\mathrm{dm}}$ increases, the number of stars having a halo structure increases. Halo stars are most interesting for our study as they can have gravitational signatures. Massive stars having abundant halo structures can significantly affect the ST in and around the NS. Fig. \ref{mr_dm} shows that as the bare mass increases (DM EoS becomes softer), the number of stars with halo structure reduces, given a fixed amount of DM fraction. Therefore, to have massive stars with a significant halo structure, one needs a relatively stiffer DM EoS along with a significant DM fraction. 

The magnetic field affects the star in a twofold way. The magnetic pressure deforms the ST in and around the star in a nonsymmetric way. The deformation not only changes the spatial structure of the star but also the total mass of the star. Although the magnetic field does not directly interact with the DM, as the ST is deformed, the DM is affected by the deformed ST. Therefore, the magnetic field has an indirect effect on the DM, which ultimately affects the structure and mass of the magnetized DMANS. 

\begin{figure}
    \centering
    \includegraphics[scale=0.445]{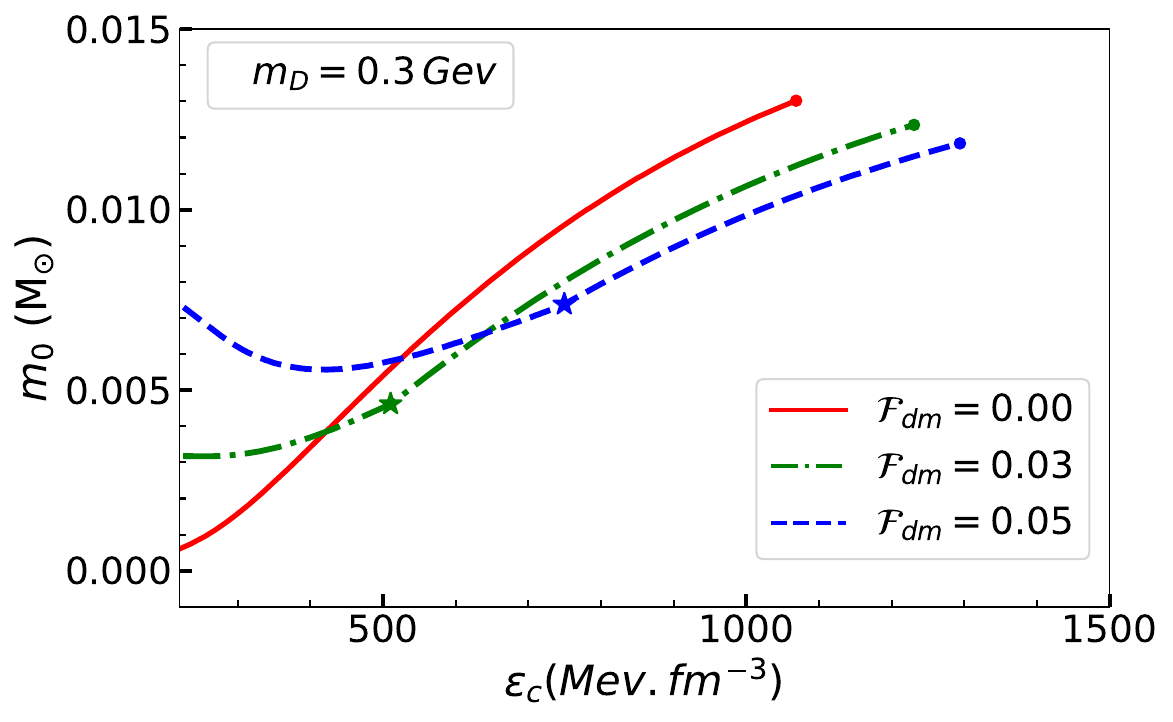}
    \includegraphics[scale=0.45]{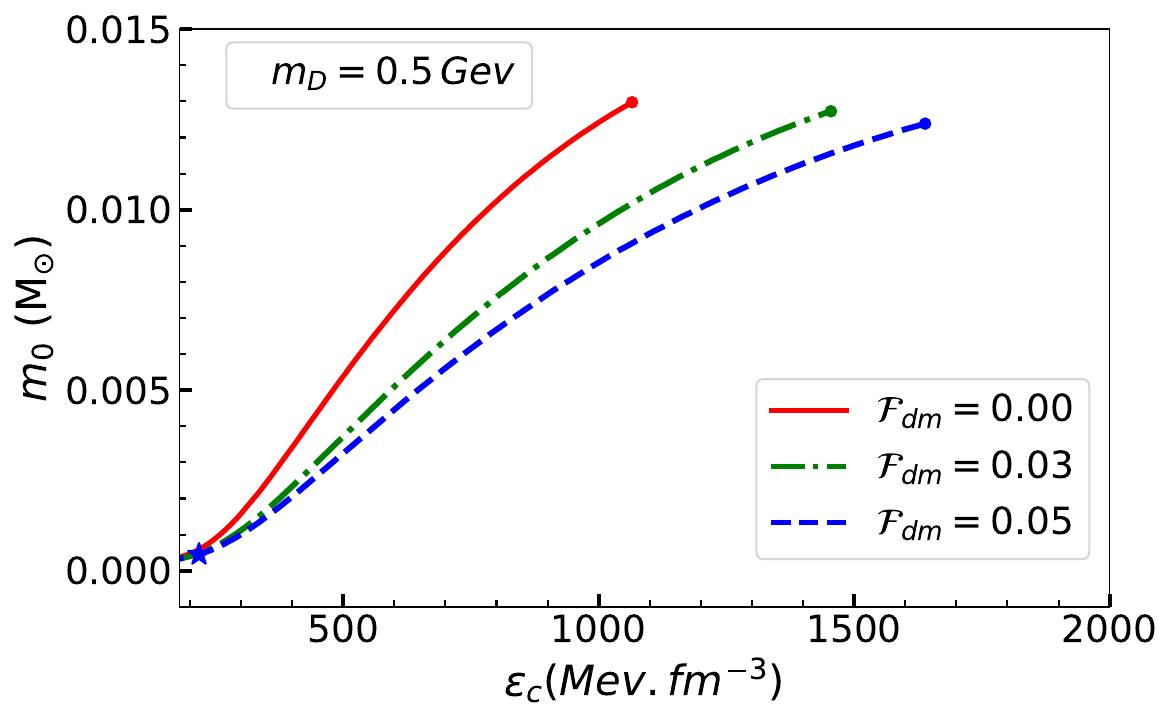}
    \caption{The variation of the additional gravitational mass ($m_0$) due to magnetic corrections in magnetized DMANSs, plotted as a function of central energy density ($\varepsilon_c$). The upper panel corresponds to a fixed dark matter bare mass of $m_D = 0.3$ GeV, and the lower panel to $m_D = 0.5$ GeV, both shown for various dark matter fractions ($\mathcal{F}_{dm}$). The red solid thick line indicates the same variation for pure magnetized HM NS (without DM, $\mathcal{F}_{dm}=0.00$), while the green dot-dashed and blue dashed lines show the same for magnetized DMANSs of DM fraction ($\mathcal{F}_{dm}$ = 0.03, 0.05). The star mark indicates that, to its right, magnetized DMANSs have a DM core, while to its left, they possess a DM halo.} 
    \label{m0}
\end{figure}

Fig. \ref{m0} shows how the magnetic field effectively influences the overall mass of the star. As the magnetic field strength increases with the baryon number density (also with central energy density $\varepsilon_c$), the additional mass of the star $m_0$ increases. For NS with no DM, there is a gradual and stiff increase of $m_0$ with the magnetic field (red-solid line of Fig. \ref{m0}). However, with the increase in DM fraction, the effective EoS (HM with DM) is softened, and the strength of the magnetic field is reduced (as it only varies with HM). Therefore, the $m_0$ curve becomes less stiff.

The DM is not totally blind to the magnetic field and responds to the curvature created by the magnetic field. Therefore, as the DM fraction increases, the initial value of $m_0$ increases at low energy density (blue-dashed and green dash-dotted curves), where the DM is much stiffer than the HM. These curves show non-monotonous behaviour as the stiffness of DM and HM changes after $\varepsilon_c \sim 300$ MeV fm$^{-3}$ (refer to Fig. \ref{eos}). For stiff DM EoS with a high fraction, initially $m_0$ is higher, and then decreases. 
However, as HM EoS starts to dominate at relatively higher densities, $m_0$ again increases, resulting in the observed non-monotonous behaviour.
However, if the DM EoS is softer (higher $m_D$ value), this effect is reduced, as can be seen in the lower panel of Fig. \ref{m0}.

\begin{figure}
    \centering
    \includegraphics[scale=0.45]{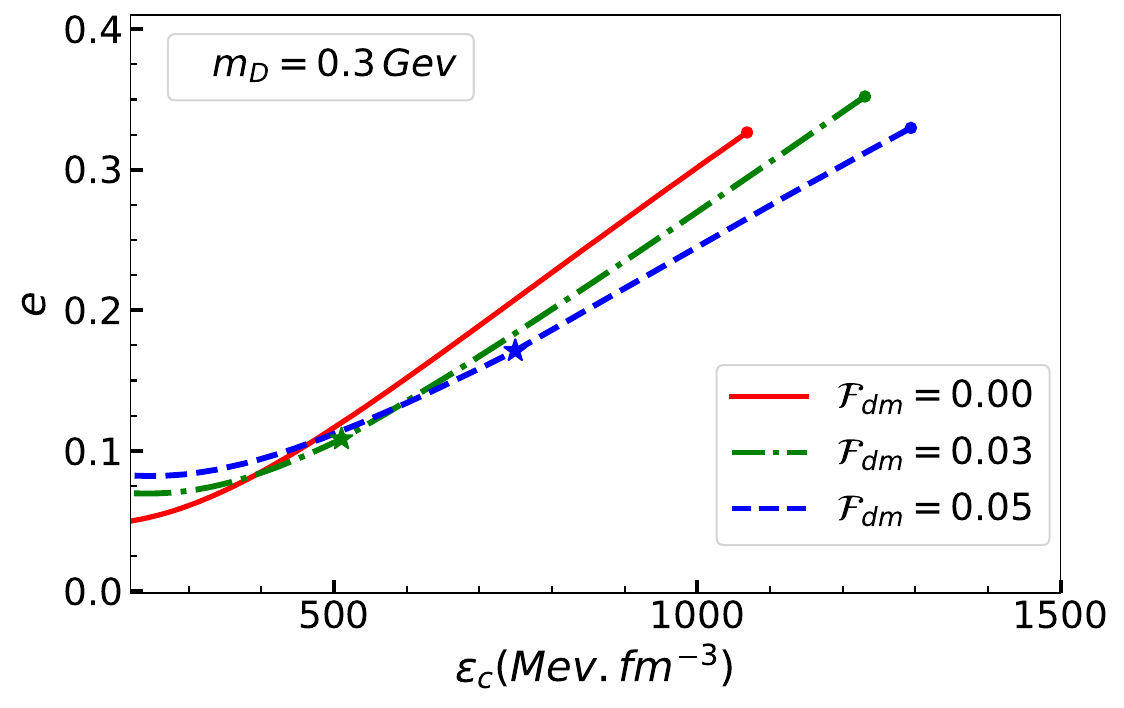}
    \includegraphics[scale=0.45]{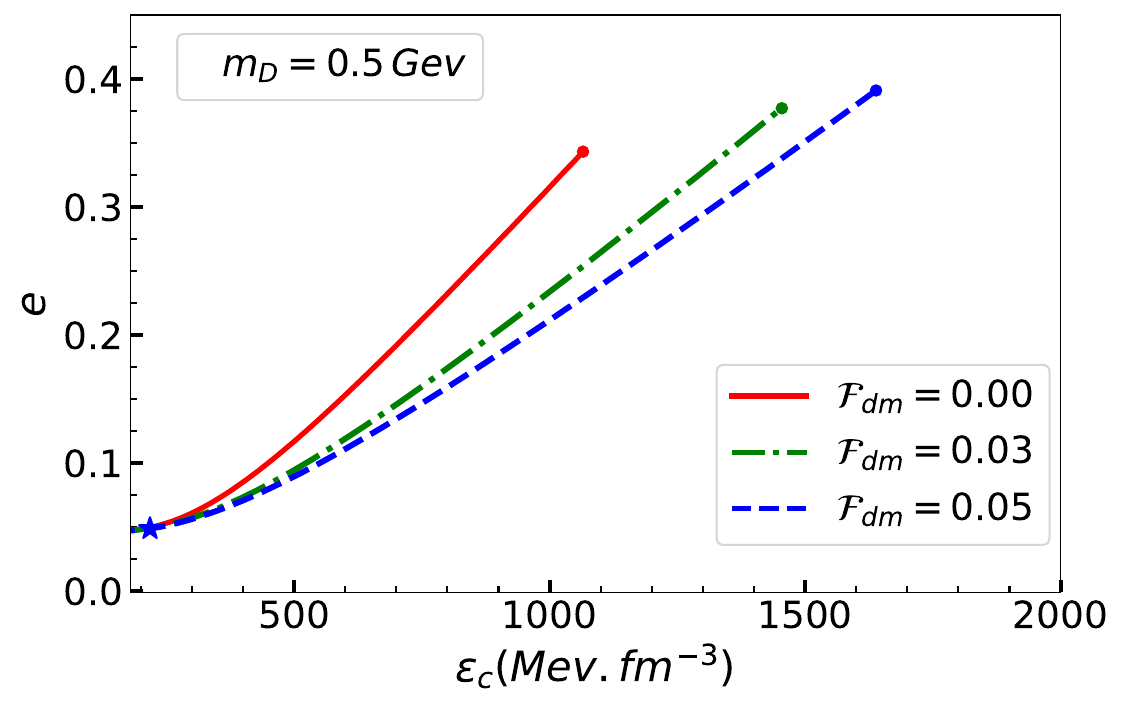}
    \caption{ Eccentricity ($e$) of magnetized DMANSs plotted against central energy density ($\varepsilon_c$). The upper panel corresponds to $m_D = 0.3$ GeV, and the lower panel to $m_D = 0.5$ GeV, with each curve representing a different dark matter fraction ($\mathcal{F}_{dm}$). The solid red curve denotes a magnetized NS without dark matter ($\mathcal{F}_{dm}=0.00$); the remaining curves correspond to magnetized DMANSs with increasing DM fraction ($\mathcal{F}_{dm}$ = 0.03, 0.05). The presence of a DM halo is indicated at lower central energy densities, to the left of the star-marker point. Beyond this point, dark matter transitions to forming a central core.}
    \label{elep}
\end{figure}

As already stated, the magnetic field not only affects the mass but also deforms the star. This can be seen in Fig. \ref{elep}, where the eccentricity (e) of the star is shown as a function of central energy density ($\varepsilon_c$). The eccentricity is defined as $ e = \sqrt{1-\left(\frac{R_p}{R_e}\right)^2}$, where \( R_e \) and \( R_p \) represent the equatorial and polar radii of the magnetized DMANS, respectively.
It is seen that as the magnetic field increases (with an increase in $\varepsilon_c$), the star becomes more elliptical or more deformed. For a pure HM NS, the eccentricity curve is very stiff. As $\mathcal{F}_{\mathrm{dm}}$ increases, the eccentricity does not increase as sharply as for a pure HM NS, as the magnetic field is only contributed by the HM. Also, a similar feature of higher eccentricity at very small densities is seen as the DM fraction is more compared to the HM at small densities. However, as the DM EoS becomes soft (high $m_D$ value, right panel) or if $\mathcal{F}_{\mathrm{dm}}$ in the star decreases, this effect decreases and the eccentricity approaches that of a pure HM star.

\begin{figure*}
    \centering
    \includegraphics[scale=0.45]{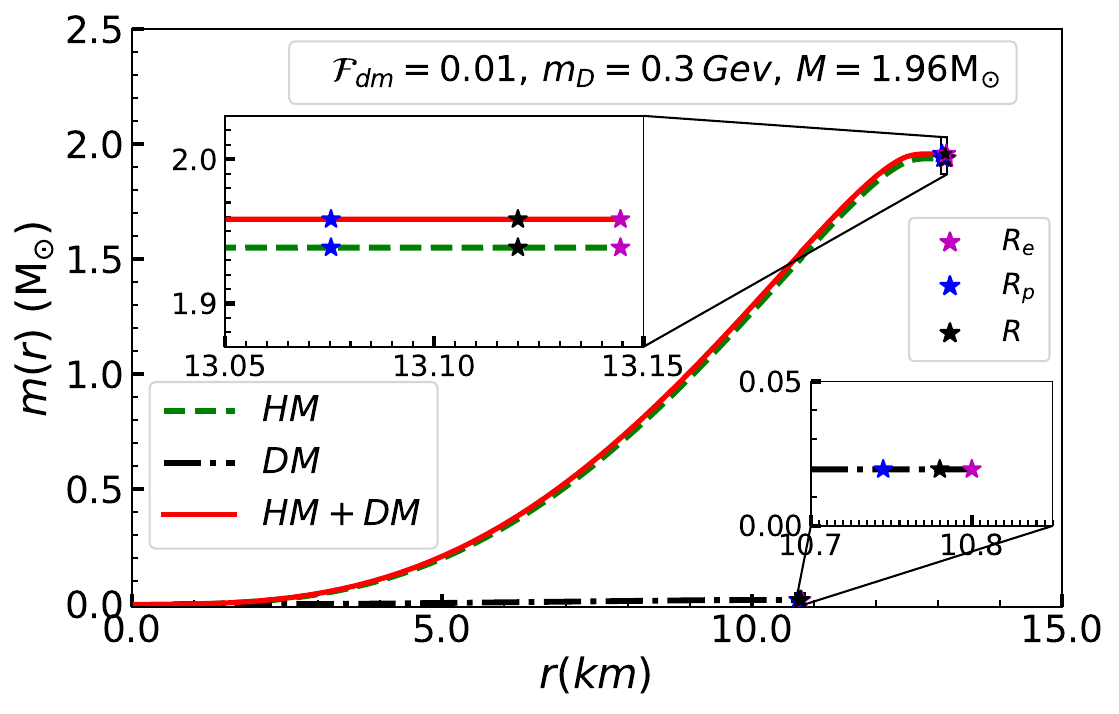}
    \includegraphics[scale=0.45]{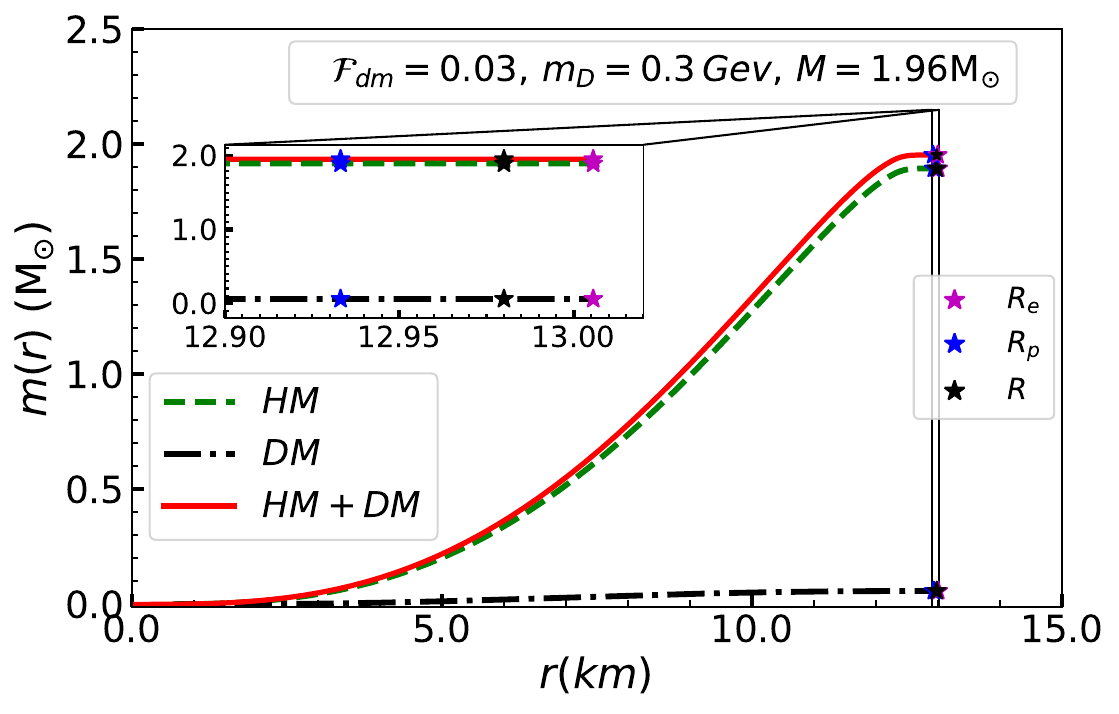}
    \includegraphics[scale=0.45]{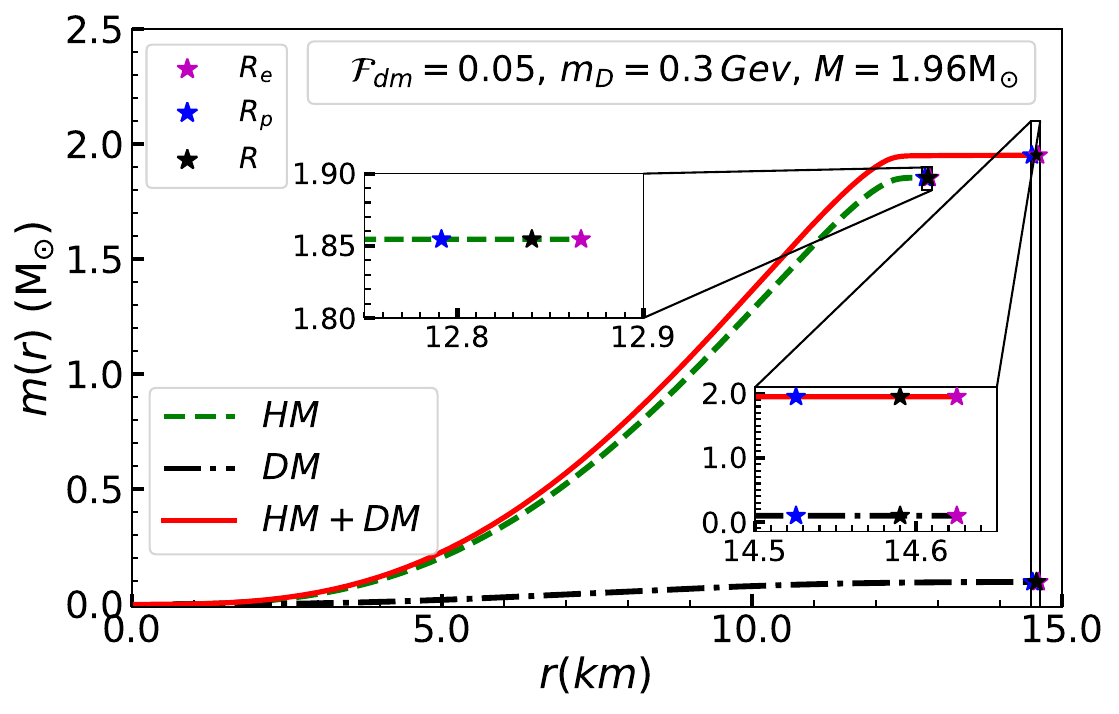}
    \includegraphics[scale=0.43]{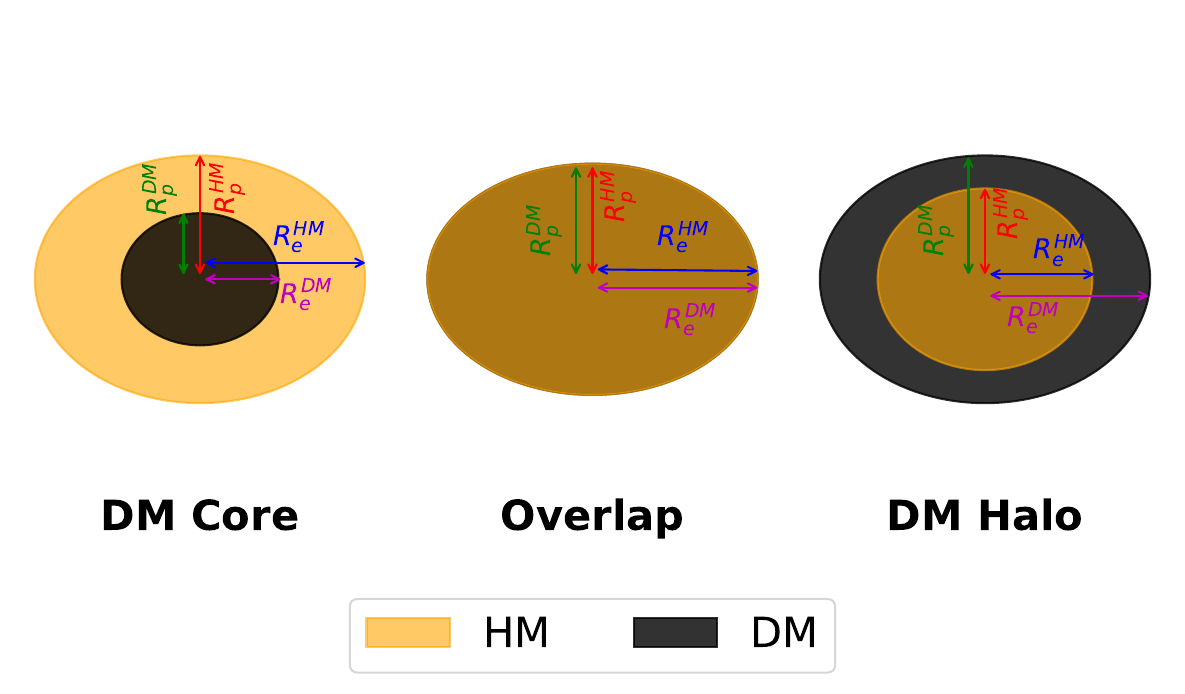}
    \caption{(\textbf{Upper left}) plot shows the mass–radius ($m(r)$-r) profile of a magnetized DMANS with a total mass of 1.96$M_\odot$, having DM fraction $\mathcal{F}_{dm} = 0.01$  and a dark matter bare mass of ($m_D = 0.3$ GeV). For this configuration, dark matter is concentrated at the core ($R_{HM}>R_{DM}$). The green dashed and black dot-dashed lines represent the M-R profiles of hadronic matter (HM) and dark matter (DM), respectively, from the center to the surface of the star. The solid red curve denotes the total M-R profile (HM + DM) of the star. The blue and magenta star markers indicate the stellar surface along the polar($R_p$) and equatorial($R_e$) directions, respectively, while the black star marker represents the surface of the corresponding non-magnetized DMANS. The inset highlights the changes in the radii of the DM, HM, and total components in the presence of a magnetic field. (\textbf{Upper right}) The plot shows the M-R profile for the same mass but different dark matter fractions $\mathcal{F}_{dm} = 0.03$, illustrating an overlapping configuration where the HM and DM have the same radius ($R_{HM}=R_{DM}$). (\textbf{Lower left}) where the DM exhibits a halo-like structure ($R_{HM}<R_{DM}$) in the star with the same mass and a dark matter fraction of $\mathcal{F}_{dm} = 0.05$. (\textbf{Lower right}) A schematic configuration showing three representative DM configurations depicted is (i) a DM core, (ii) an overlapping configuration in which DM and HM share the same polar and equatorial radii, and (iii) a DM halo.}
    \label{mr-1.9}
\end{figure*}

The impact of non-symmetric deformation induced by the magnetic field in magnetized DMANS becomes evident from the M–R curve plotted for isolated stars. As argued previously, the effect is more pronounced when we have a significantly massive NS (stars above $1.4\,M_\odot$ ). As the effect of DM is more pronounced if the EoS of DM is stiff and has a significant $\mathcal{F}_{\mathrm{dm}}$ value in the star, we restrict our analysis to a minimum bare mass $m_D=0.3$ GeV.
In Fig. \ref{mr-1.9} we show how the mass ($m$) varies with radial distance ($r$) for magnetized DMANS with mass \( M = 1.96\,M_\odot \). As expected for a massive star, if $\mathcal{F}_{\mathrm{dm}}$ is small ($\mathcal{F}_{dm} = 0.01$), the DM is only restricted inside the star, and at the surface we only have HM. Therefore, its electromagnetic and gravitational signatures (which originate from the surface) are governed by the HM. Although the total M-R structure is governed by both HM and DM, there is no way to infer DM properties for such stars. 
The magnetic field deforms the star more along the polar direction than along the equatorial direction. It is also evident from the plot that, since HM resides more toward the outer regions, it experiences greater deformation compared to DM. This effect becomes more pronounced for more massive stars, where the dark matter is concentrated deeper inside the star. This behavior is also clearly observed in Fig.~\ref{mr}, where stars with higher mass exhibit greater overall deformation. In the upper right panel of Fig.~\ref{mr-1.9}, we present a star of the same mass with DM fraction ($\mathcal{F}_{dm} = 0.03$) in which the radial extent of DM is comparable to that of HM. This configuration exhibits similar deformation along both the equatorial and polar directions.

An interesting feature emerges in the lower left panel of Fig.~\ref{mr-1.9}, where the DM fraction is higher ($\mathcal{F}_{dm} = 0.05$) for a star of the same mass. In this case, the DM forms an extended halo beyond the entire magnetized DMANS. The deformation along both the equatorial and polar directions is more pronounced compared to that observed in the HM case. Specifically, the deformation along the polar direction can reach several hundred meters, and this effect becomes more pronounced for stars with higher mass and a larger DM radius. Such a DM halo can have a significant impact on both the electromagnetic and gravitational signatures of the NS. Light coming from any stars behind the magnetized DMANS can have a lensing effect due to the halo. Furthermore, the effect would be non-symmetric, which is expected to have very distinct signatures. 
The non-symmetric halo is only possible in magnetars as the magnetic field in them is a few orders greater than any other system found in the cosmos. The lower right panel of Fig.~\ref{mr-1.9} shows an illustrative configuration corresponding to the three DM configurations described above.

\begin{figure}
    \centering
    \includegraphics[scale=0.45]{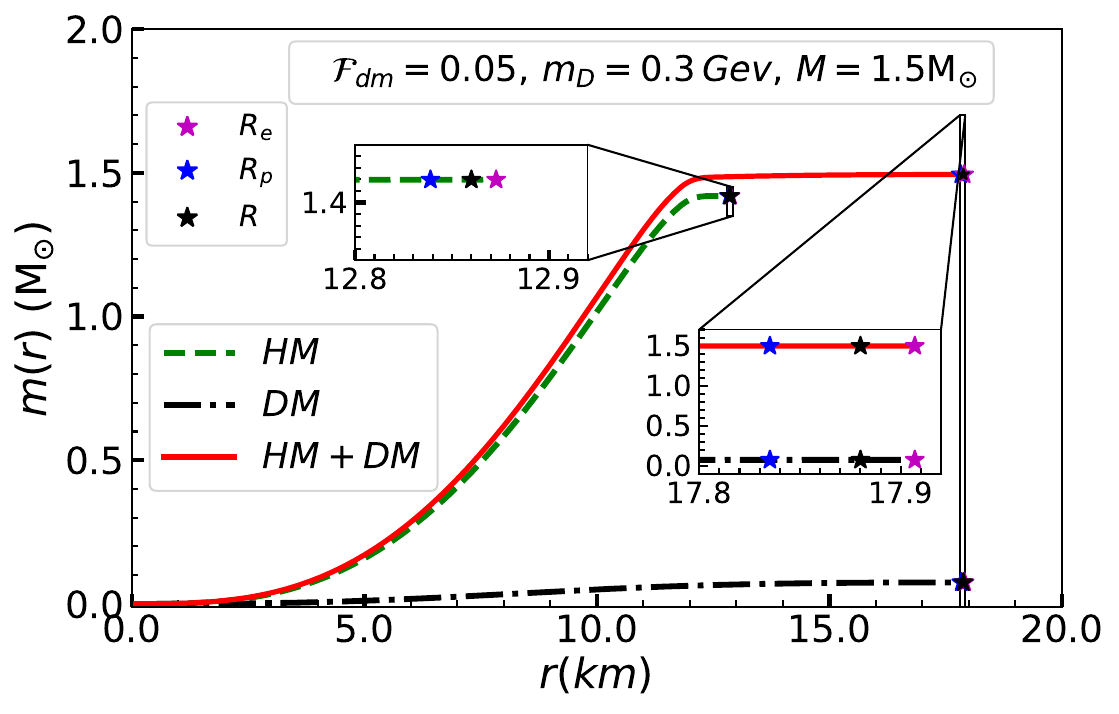}
    \includegraphics[scale=0.45]{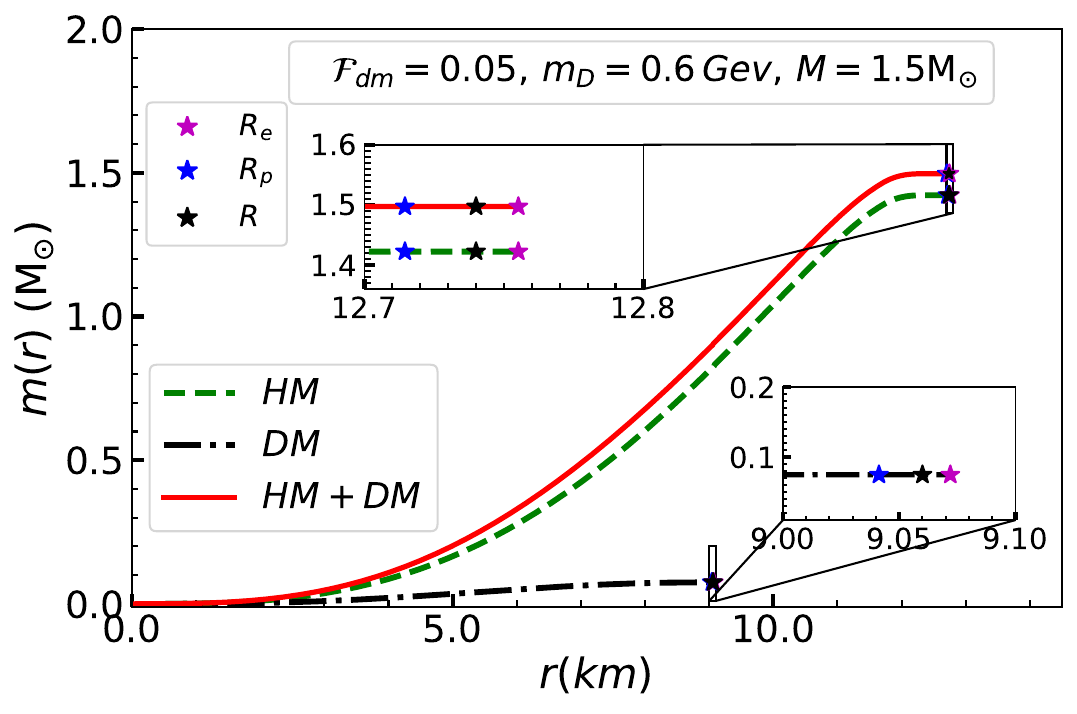}
    \caption{ The $m(r)$ vs. $r$ profile for a $1.5~M_\odot$ star with a DM fraction of $\mathcal{F}_{dm} = 0.05$ shows that, for a DM bare mass of $m_D = 0.3$ GeV, the dark matter forms a halo-like structure (\textbf{upper panel}), while for a bare mass of $m_D = 0.6$ GeV, the same star exhibits dark matter concentrated at the core (\textbf{lower panel}). In the inset, the M-R profile is shown to illustrate the extent of deformation.}
    \label{mr-1.5}
\end{figure}

In Fig.~\ref{mr-1.5}, We present the M–R profiles for two stars with identical total masses of \(1.5\,M_\odot\) and same DM fraction ($\mathcal{F}_{dm} = 0.05$), but different values of the \(m_D\), illustrating the effect of the stiffness of the DM EoS. 
In the upper panel of Fig.~\ref{mr-1.5}, corresponding to a relatively stiff DM EoS ( \(m_D = 0.3\,\mathrm{GeV}\)), the outer structure of the magnetized DMANS is dominated by an extended DM halo, while the HM remains confined to the core. The deformation in the DM component is significantly larger than that in the HM. In this specific case, the polar and equatorial radii of the DM component change by approximately $45\,\text{meters}$ and 30 meters from sphericity, respectively, whereas for the HM, the corresponding deformations are about 20 meters and 15 meters. The gravitational signature of such a star is primarily determined by the non-symmetric DM halo.
As the DM EoS becomes softer ($m_D = 0.6 \mathrm{GeV}$), the halo structure disappears, even for a star with the same total mass and the same amount of dark matter ($\mathcal{F}_{dm} = 0.05$), as shown in the lower panel of Fig.~\ref{mr-1.5}. In this case, the deformation induced by the magnetic field is more prominent in the HM compared to the DM. The difference between the equatorial and polar radii of the HM is approximately 50 meters. For such a configuration, the observational signature is primarily governed by the HM distribution rather than by the DM.
\begin{figure}
    \centering
    \includegraphics[scale=0.45]{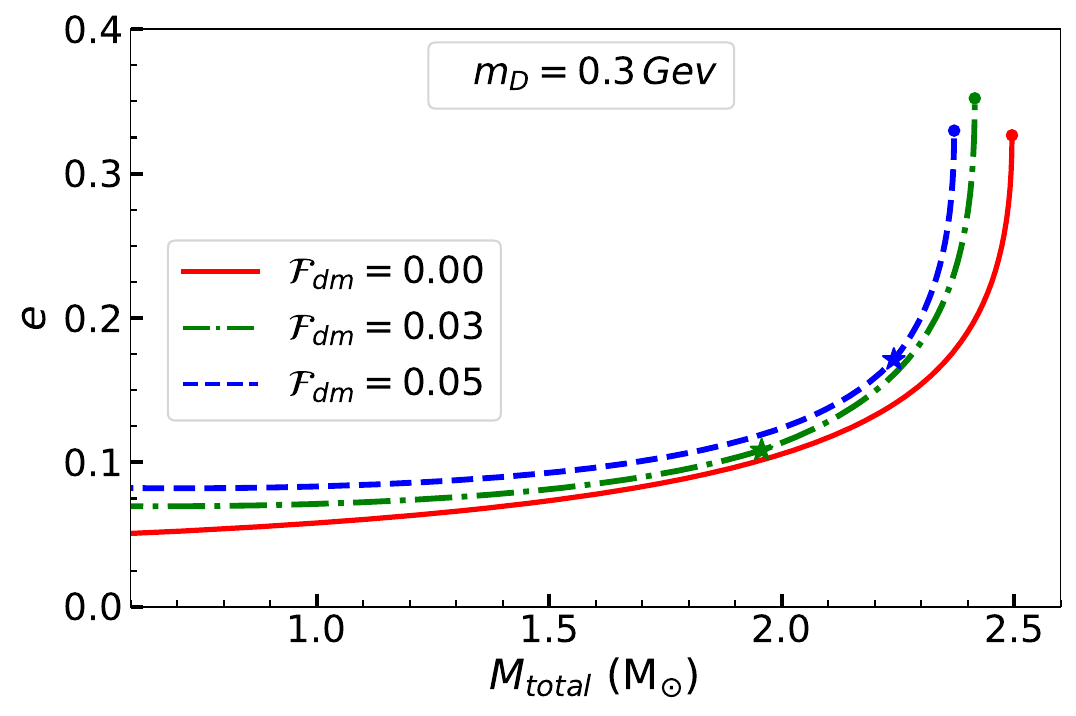}
    \caption{The eccentricity ($e$) is plotted as a function of the total mass for magnetized DMANSs with $m_D = 0.3$ GeV, considering varying $\mathcal{F}_{\mathrm{dm}}$. The star marks the transition point where the nature of the DM distribution changes from a halo-dominated configuration on the left to a core-dominated one on the right.
    }
 \label{em_for_0.4}
\end{figure}

As evident from the $M$-$R$ relation, more massive stars experience greater deformation due to the presence of a higher magnetic field. To quantify this effect across different stellar masses, we present in Fig.~\ref{em_for_0.4} the variation of eccentricity as a function of mass for magnetized DMANSs with \(m_D = 0.3\,\mathrm{GeV}\) and varying DM fractions. We observe that eccentricity increases with mass, although the growth is initially slow. Beyond approximately \(1.5\,M_\odot\), the deformation becomes significantly larger. The star marker in the figure denotes the configuration where the DM and HM have equal radii and exhibit identical deformation. Furthermore, for a given mass, increasing the DM fraction results in greater overall deformation of the star.

The presence of a DM halo around NSs can potentially be confirmed using existing observational methods; however, the precision of these measurements must be significantly improved. One promising approach involves identifying discrepancies in the measurement of continuous gravitational wave (GW) amplitudes emitted by NSs.
Another way would be the discrepancy in the observed and estimated period and period change of an isolated magnetar having a halo.

\begin{figure}
    \centering
    \includegraphics[scale=0.45]{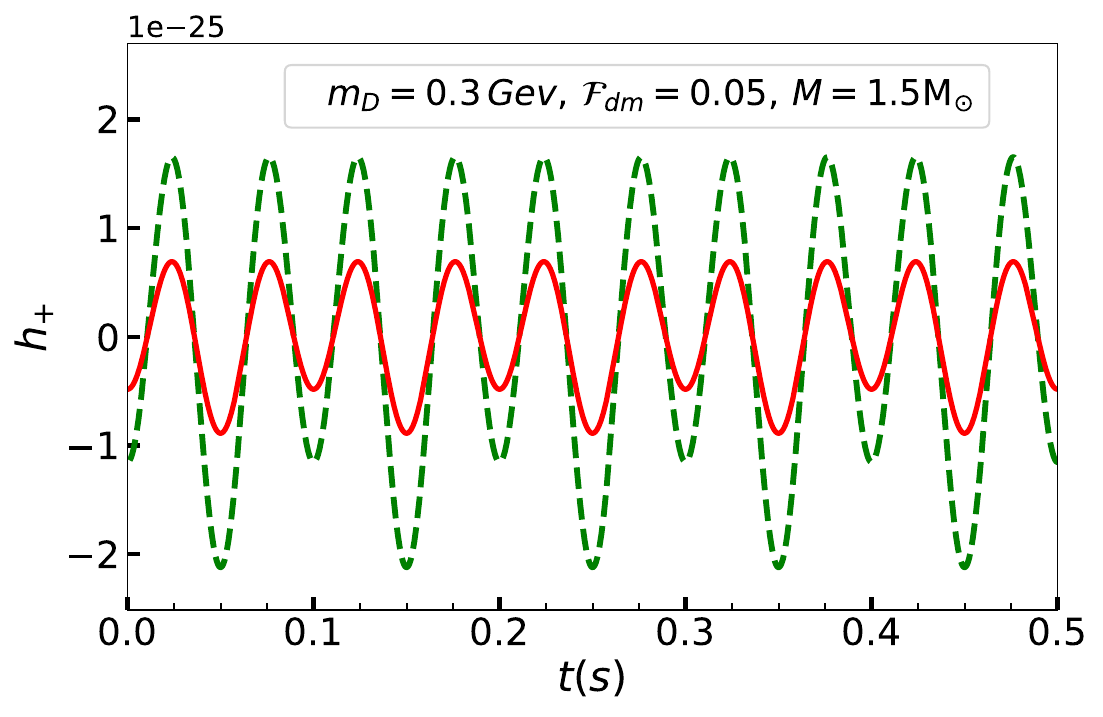}
    \caption{The plot of continuous GW amplitude against time for a slowly rotating, magnetically deformed star in the shape of an oblate spheroid. The red-coloured solid line denotes GW for magnetized DMANS considering the observed radius (HM radius). The green-coloured dashed line shows the expected GW for magnetized DMANS with a DM halo.}
 \label{Gw amp}
\end{figure}

Although rotation is not considered in the present study, it is well established that NSs rotate about their axes. Among them, NSs possessing extremely strong magnetic fields, known as magnetars, are generally characterized by slow rotation compared to typical radio pulsars. In the case of slow rotation, rotational deformation of the star is expected to be negligible. However, the strong internal magnetic field can induce significant deformation, leading to a non-axisymmetric distribution.

Assuming the magnetic axis is misaligned with the rotational axis, a condition commonly invoked to explain the pulsar phenomenon, a magnetically deformed, slowly rotating neutron star can emit continuous gravitational waves (GW). This emission arises due to the time-dependent mass quadrupole moment of the star. The leading contribution to the gravitational radiation field is then governed by the quadrupole formula given as \citep{Zimmermann:1979ip,Bonazzola:1995rb,Jaranowski:1998qm,Sieniawska:2019hmd}:
\begin{align}
h_{+} &= h_0 \sin \psi \left[ \frac{1}{2} \cos \psi \sin i \cos i \cos \Omega t \right. \nonumber \\
&\hspace{7.4em}\left. \quad - \sin \psi \frac{1 + \cos^2 i}{2} \cos 2\Omega t \right] \\
h_{\times} &= h_0 \sin \psi \left[ \frac{1}{2} \cos \psi \sin i \sin \Omega t - \sin \psi \cos i \sin 2\Omega t \right]
\end{align}
Where
\begin{align}
   h_0=\frac{4G}{c^4}\frac{\Omega^2 I \varepsilon}{d}  
\end{align}
$h_0$ is the GW amplitude, $d$ is the distance of the magnetar, $c$ the speed of light, ($\Omega = 2\pi/P$) is the rotational velocity of the magnetar, $I$ the moment of inertia with respect to the rotational axis, $\varepsilon$ is the ellipticity of the star, and $\psi$ is the angle between the rotation and magnetic axis, $i$ denotes the line-of-sight angle. The ellipticity of the star is geometrically related to its eccentricity and can be expressed as 
\begin{IEEEeqnarray}{rCl}
    \varepsilon = \frac{1}{2}e^2
\end{IEEEeqnarray}
Due to the distinct responses of DM and HM to the magnetic field, their respective deformations differ. This asymmetry in deformation manifests as a difference in the resulting GW amplitudes.   
Therefore, an estimate can be made of the difference in the GW amplitude arising from the discrepancy in eccentricity between the visible star and the star including the DM halo. For the same star, the observed eccentricity (i.e., the HM eccentricity, $e_{hm}$) differs from the intrinsic eccentricity of the star when the DM halo is taken into account.
Fig.~\ref{Gw amp} shows the variation of GW for magnetized DMANS of mass $1.5\,M_\odot$,with a dark matter bare mass of $m_D = 0.3$ $\mathrm{GeV}$ and a dark matter fraction $\mathcal{F}_{dm} = 0.05$. The different colored lines show the effect of the DM halo on continuous GWs. The solid green line represents the expected GW signal considering the DM halo, while the solid red line corresponds to the GW amplitude computed using only the observed HM radius. To compute the GW amplitude for the $1.5\,M_\odot$ magnetized DMANS having radius $R_{e} = 17.914\,\text{km}$ and $R_{p} = 17.837\,\text{km}$ ($R_e^{hm} = 12.876\,\text{km}$ and $R_p^{hm} = 12.832\,\text{km}$), we consider the distance of observation to be $d = 10$ \text{kpc}. We assumed the misalignment angle ($\psi$) about $\pi/6$, the inclination angle ($i$) to be $\pi/9$ and the rotation period ($P$) to be $0.1\,\text{s}$ ($\Omega \approx 62.83 \,\text{rad/s}$). 
The moment of inertia ($I$) of the star is given by $I = \frac{2}{5}MR_e^{2}$. With these values, the amplitude of the GW ($h_0$) is calculated to be $2.909 \times 10^{-25}$. However, when the DM halo is taken into account, the expected amplitude increases to $6.951 \times 10^{-25}$. Given the capabilities of the advanced VIRGO detector \citep{VIRGO:2014yos} and the advanced LIGO detector \citep{LIGOScientific:2014pky}, it may be possible to detect gravitational waves of this order of amplitude.

We can also make some estimations of the GW strength for some well-known pulsars whose masses are still unknown. Assuming these stars have a mass of approximately $1.5\,M_\odot$, we can compute their moment of inertia ($I$) and ellipticity ($\varepsilon$) from our calculations.
For instance, considering the magnetar SGR 1806$-$20, which has a surface magnetic field of the order of $10^{15}\,\text{G}$, a pulse period of $P = 7.56\,\text{s}$, and is located at a distance of approximately $d \approx 8.7\,\text{kpc}$, the calculated GW amplitude is $h_0 = 5.851 \times 10^{-29}$. However, if we consider this magnetar to possess a DM halo, the value of $h_0$ changes to $1.398 \times 10^{-28}$. A similar estimation for the magnetar SGR 0501+4516, which has a surface magnetic field of approximately $10^{14}\,\text{G}$, a pulse period of $P = 5.76\,\text{s}$, and is located at $d = 2\,\text{kpc}$, yields $h_0 = 4.384 \times 10^{-28}$. Therefore, the maximum GW amplitudes from these magnetars are likely of the order of $10^{-28}$, which remains undetectable by current GW detectors. Although these values are beyond the capability of present-day detectors, the values clearly show the difference accounted for the halo.

Another potential method for detecting DM halo around a magnetar is through discrepancies in the estimation of the spin period and its time derivative in isolated magnetars. Assuming a vacuum dipole model for the magnetar, with the exterior dipole field inclined at an angle $\psi$ with the rotation axis, the rotational period ($P$) and rate of change of NS's period ($\dot{P}$) are given by \citep{2001ARep...45..389M,Ridley:2010vr}

\begin{equation}
P=\sqrt{P_{0}^{2}+2 A t},\label{period}
\end{equation}

\begin{equation}
\dot{P}=\frac{A}{\sqrt{P_{0}^{2}+2At}},\label{period derivative}
\end{equation}

Where $P_0$ is the initial spin period at birth, $t$ is the age of the magnetar, and $A = \frac{8 \pi^{2} B_{s}^{2} R^{6}}{3 c^{3} I}$. Clearly, the quantities $P$, $\dot{P}$, and the $\dot{P}$–$P$ diagram depend on the values of the surface magnetic field ($B_s$), radius ($R$), and moment of inertia ($I$). These parameters are expected to differ if the NS is surrounded by a DM halo. The observed visible radius of the star corresponds to the HM radius, while the actual radius may extend further due to the presence of the DM halo. The difference between these two radii can be of the order of hundreds to thousands of meters, leading to discrepancies between the calculated values of $P$ and $\dot{P}$ when using the actual radius (including the DM halo) versus the radius inferred from electromagnetic observations (i.e., the visible HM radius $R_{hm}$). Such discrepancies will eventually affect the $\dot{P}$–$P$ diagram, the characteristic age ($\tau = \frac{P}{2\dot{P}}$) \citep{2001ARep...45..389M}, the spin-down luminosity ($L_{\text{sd}} = \frac{4\pi^2 I \dot{P}}{P^3}$) \citep{Possenti:2001uu}, and the inferred magnetic field evolution of the pulsar \citep{1977ApJ...215..302F,Igoshev:2021ewx}. Assuming a $1.5,M_\odot$ magnetar with a dark matter fraction ($\mathcal{F}_{dm} = 0.05$), a dark matter bare mass ($m_D = 0.3\,\mathrm{GeV}$), a surface magnetic field of the order of $10^{15}\text{G}$, and an initial rotation period of $P_0 = 5,\text{s}$, the period and its derivative after $10^6$ years are calculated using Eqs.~(\ref{period}) and (\ref{period derivative}). For the case considering only the observed HM radius, the period is found to be $P = 5.00449\text{s}$ with a period derivative of $\dot{P} = 4.488 \times 10^{-9}$. In contrast, when including the DM halo around the magnetar, the period becomes $P = 5.01680\text{s}$ and the period derivative increases to $\dot{P} = 1.677 \times 10^{-8}$. This indicates that estimations of the characteristic age ($\tau$), spin-down luminosity ($L_{\text{sd}}$), and magnetic field evolution will differ significantly depending on whether the DM halo is considered in the NS model.

To assess the validity of the perturbative expansion in the regime of extreme magnetic fields for magnetized DMANS, we evaluate both global and local measures. The global ratio of magnetic to gravitational binding energies is defined as \(\eta_{B/G} = E_{\text{mag}} / E_{\text{grav}}\), where the magnetic binding energy is given by \(E_{\text{mag}} = B_s^2 R^3 / 6\), with \(B_s\) denoting the surface magnetic field and \(R\) the stellar radius. The gravitational binding energy for a magnetized DMANS can be expressed as \(E_{\text{grav}} = \tfrac{0.6 \beta}{1 - 0.5 \beta} M\) \citep{Lattimer:2000nx}, where \(\beta = M/R\) is the compactness parameter. 
For a \(1.5\,M_\odot\) star with a surface magnetic field \(B_s = 10^{15}\,\text{G}\), a dark matter fraction of \(\mathcal{F}_{dm} = 0.05\), a dark matter bare mass of \(m_D = 0.3\,\mathrm{GeV}\), and a stellar radius \(R = R_e = 17.914\,\text{km}\), the ratio of magnetic to gravitational binding energy becomes \(\eta_{B/G} \approx 4.514 \times 10^{-6}\), which clearly lies well within the perturbative regime (\(\eta_{B/G} \ll 1\)). Even for the maximum-mass (\(2.495\,M_\odot\)) magnetized star without dark matter, we obtain \(\eta_{B/G} \approx 3.223 \times 10^{-7} \ll 1\), further confirming the applicability of the perturbative expansion. 
As a complementary local check, we calculate the ratio of magnetic to matter pressure at the stellar center, defined as \(\chi_{B/P} = P_B / P_{\mathrm{matter}} = B_c^2 / (8\pi P_c)\), where \(B_c\) and \(P_c\) are the central magnetic field and central matter pressure, respectively. For a \(1.5\,M_\odot\) star we obtain \(\chi_{B/P} \approx 2.74 \times 10^{-2} \ll 1\), while for a \(2.495\,M_\odot\) star the value is \(\chi_{B/P} \approx 6.23 \times 10^{-2} \ll 1\). For both stars, the magnetic pressure remains of the order of $10^{-2} - 10^{-1}$ relative to the matter pressure up to near the stellar surface.
Therefore, both global and local estimates consistently demonstrate that the perturbative expansion remains well justified across the parameter space explored.

\section{Summary and discussion}\label{sum_disc}
 In this work, we have studied the effect of the magnetic field on the structure of DMANS. We have assumed that the HM is significantly stiffer than DM at high densities and have varied the stiffness of the DM by a single variable, $m_D$. We investigate the effects of magnetic fields for dark matter bare mass values ranging from $m_D = 0.3$ to $0.6\,\mathrm{GeV}$, and dark matter fraction $\mathcal{F}_{\mathrm{dm}}$ up to $5\%$. The HM is kept constant. The poloidal magnetic field is varied as a function of baryonic density, and it increases with an increase in baryonic density. The surface field is kept constant at $10^{15}$ G. Such a high magnetic field is assumed to have significant deformation in the star. 

It is found that the magnetic field deforms the star and is responsible for the bulging of the equator and compression of the poles. The presence of DM makes stars more compact, and a higher DM fraction with a stiff DM EoS causes a DM halo around the HM. The deformation of the star reduces with the increase in DM fraction as the magnetic field strength goes down, and the fraction of HM reduces. It is also seen that the softer the DM EoS, the more the deformation of the star. The magnetic field is also responsible for making the star more massive. However, as $\mathcal{F}_{\mathrm{dm}}$ increases, the additional mass due to magnetic deformation decreases as the mass contribution due to HM reduces significantly.

The magnetic field influences DMANS in a distinctive manner. For configurations with a significant DM fraction, a non-symmetric DM halo can form outside the stellar surface. This halo exhibits an equatorial bulge and polar compression, resulting in a non-trivial, asymmetric structure. Such a scenario is likely to arise in magnetars, where extremely strong magnetic fields are present. The asymmetry between the equatorial and polar regions of the DM halo can lead to differing gravitational imprints. There would be a discrepancy in the period and period derivative of the star if the halo is not considered. Above this, the continuous gravitational wave amplitude (and even the frequency) would also differ from that if the halo is not considered. 
It is to be understood that rotation can also generate deformation in the NS structure. However, the case with magnetic deformation is simply unique. Rotation cannot generate deformation as generated by a magnetic field, as it affects both the DM and HM similarly. Although DM does not directly interact with the magnetic field, it is affected by the magnetic field distortion of ST. The DM responds to this ST deformation gravitationally and leaves a unique imprint.

The results obtained are very robust and do not depend on the microphysics of the HM and DM EoS. As long as DM EoS is softer than HM EoS, the qualitative aspect of our results does not change. The only other aspect that needed to have the same qualitative results is the strength of the magnetic field. The magnetic field needs to be high to produce significant deformation in DMANS such that one has measurable observational signatures. This can happen only in magnetars. Although the results obtained are for a simple case of both DM and magnetic field configuration, one can improve upon this to have a more detailed and accurate study involving numerical and computational studies of magnetic field configuration. 
The presence of only the toroidal magnetic field component induces prolate deformation in neutron stars. In dark matter admixed systems, such deformations are expected to affect the surrounding halo, rendering it non-symmetric. However, toroidal fields are difficult to model analytically and are difficult to consistently treat within perturbative frameworks \citep{Kiuchi:2008ss}. More realistic and stable configurations, such as mixed poloidal--toroidal fields, tilted torus, or twisted-torus structures, introduce competing deformations that further complicate predictions. Accurately capturing these effects requires fully relativistic magnetohydrodynamic (MHD) simulations or other numerical approaches \citep{ Lasky:2013bpa,Mastrano:2015rfa,Tsokaros:2021pkh}. In this study, we focus solely on the poloidal magnetic field. In future work, we aim to extend our analysis to include toroidal and mixed field configurations using numerical codes within a two-fluid formalism, enabling self-consistent modelling of magnetic deformation and its impact on the dark matter halo.

\section*{Acknowledgements}

The authors AK and RM thank the Indian Institute of Science Education and Research Bhopal for providing all the research and
infrastructure facilities. RM acknowledges the Science and Engineering Research Board (SERB), Govt. of India, for financial support in the form of a Core Research Grant (CRG/2022/000663).

\section*{Data availability statement}

This is a theoretical work, and has no additional data with it.
\bibliographystyle{mnras}
\bibliography{refs}
\bsp	
\label{lastpage}
\end{document}